# Combinatorial materials discovery strategy for high entropy alloy electrocatalysts using deposition source permutations

Lars Banko[1,#], Olga A. Krysiak[2,#], Bin Xiao[1], Tobias Löffler[1,2,3], Alan Savan[1], Jack Kirk Pedersen[4], Jan Rossmeisl[4], Wolfgang Schuhmann[2], Alfred Ludwig[1,3,*]

1. Materials Discovery and Interfaces, Institute for Materials, Ruhr-University Bochum, 44801 Bochum, Germany
2. Analytical Chemistry – Center for Electrochemical Sciences (CES), Faculty of Chemistry and Biochemistry, Ruhr University Bochum, Universitätsstr. 150, D-44780 Bochum, Germany
3. Center for Interface-Dominated High Performance Materials, Ruhr-University Bochum, 44801 Bochum, Germany
4. Theoretical Catalysis – Center for High Entropy Alloy Catalysis (CHEAC), Department of Chemistry, University of Copenhagen, Universitetsparken 5, 2100 Copenhagen, Kbh (Denmark)

[#] equal contributions

[*] corresponding author

**Abstract**

**High entropy alloys offer a huge search space for new electrocatalysts. Searching for a global property maximum in one quinary system could require, depending on compositional resolution, the synthesis of up to $10^6$ samples which is impossible using conventional approaches. Co-sputtered materials libraries address this challenge by synthesis of controlled composition gradients of each element. However, even such a materials library covers less than 1% of the composition space of a quinary system. We present a new strategy using deposition source permutations optimized for highest improvement of the covered new compositions. Using this approach, the composition space can be sampled in different subsections allowing identification of the contribution of individual elements and their combinations on electrochemical activity. Unsupervised machine learning reveals that electrochemical activity is governed by the complex interplay of chemical and structural factors. Out of 2394 measured compositions, a new highly active composition for the oxygen reduction reaction around $Ru_{17}Rh_5Pd_{19}Ir_{29}Pt_{30}$ was identified.**

# Introduction

The need for new materials is driven by the increasing demand for materials which enable new applications, e.g. in efficient energy conversion to combat climate change. The chemical complexity of materials used in applications is usually high, as many elements (4 to 12) are needed to adjust materials properties to meet frequently contradicting demands. Traditional examples of such materials are steels, superalloys or metallic glasses, whereas since several years new types of chemically complex materials are emerging such as high entropy alloys (HEA) or complex solid solutions (CSS)[1]. However, the poly-elemental nature of these materials makes the identification of optimal compositions for specific properties a very challenging task. CSS were identified recently as a discovery platform for novel electrocatalysts[2,3], offering an immense multidimensional search space for finding materials with enhanced properties and combinations of properties such as high activity, selectivity and stability for a given catalytic reaction by choice of constituent elements and their relative chemical composition. CSS were already successfully applied to hydrogen[4–7] and oxygen evolution reactions[6,8–10], $CO$[11,12], $CO_2$[11,13] and $O_2$ reduction reactions[2,3,12,14–16], methanol[7,17,18] and ethanol oxidation[19] as well as ammonia synthesis[20] and decomposition[21,22]. The special properties of CCS arise from their unique multi-element active site arrangements induced by the presence of a multinary single solid solution phase. The presence of at least five different elements favour the CSS formation with properties which are usually not accessible for their subsystems (i.e. quaternary, ternary, binary and unary)[23]. The evolving new opportunities create, however, a multidimensional challenge in finding a promising choice of quinary systems and optimizing the composition within each system. Particularly, 10,256 different compositions exist for only one quinary system (including all subsystems), even when only 5 at.% steps are considered. For 1 at.% steps the number of different compositions increases to $> 4*10^6$ for a single quinary system. Composition changes affect the intrinsic activity of CSS by altering active site configuration probabilities and more importantly by defining the amount of relevant active sites at the surface[24]. Therefore, compositional optimization denotes a crucial step with high impact on the discovery of new catalysts, yet it is usually not considered in literature since it is very challenging to be accomplished[25].

This challenge can be addressed by a prediction-experimental verification cycle[14], where in an iterative cycle a theoretical model is refined by experimental data from thin film materials libraries (MLs) to predict composition-activity trends in a CSS system. The complexity of the interactions translates to the assumptions made in the model, which ultimately requires multiple iterations to yield an accurate model for the experimentally covered composition range. However, as demonstrated below, each ML still covers only a very limited part of the quinary composition space and thus, the model is only aligned to a relatively small compositional region of a CSS system. While all combinations of three independent deposition sources can be projected on a planar substrate[26], the projection of five individual deposition sources covers only a subsection of all possible combinations. Hence, co-deposition from five deposition sources equipped with different elements does not necessarily provide information about the entire composition activity trend. Due to mutual information in compositional gradients, a single ML does not allow definitive conclusions to be drawn about the fundamental effects of individual elements or their combinations (SI1).

We propose to address this challenge by a new exploration strategy which effectively extends the experimentally covered composition space in a quinary system. This is based on designing a set of MLs with permutated deposition source arrangements. With each permutation of the source arrangement, each resulting ML contains a different subspace of the quinary composition space. Thus, we are able to study the contribution of individual elements and their combinations on electrochemical activity – unbiased by the confounded information from correlated compositional gradients. We show the effectiveness of this approach for CSS electrocatalyst discovery.

## Results and discussion

Compositional exploration of a quinary system by systematic permutations of five co-deposition sources

Generally, co-sputtered MLs allow covering extended ranges of multinary material systems in a single experiment in the form of continuous composition gradients. However, the composition ranges are only covered partially, and the compositional coverage of a system decreases strongly with increasing number of constituents. Figure 1A shows typical ternary and quaternary composition spreads of MLs co-sputtered from three or four sources, respectively. For quinary MLs a co-sputter arrangement and the resulting composition gradients are visualized for 342 measurement areas (MAs) over the substrate in Figure 1B. This ML was synthesized with the deposition rate of each co-deposition source adjusted to achieve an equiatomic composition in the center of the ML. Chemical compositions comprised in this quinary ML are measured using high-throughput EDX on a regular 4.5 mm x 4.5 mm measurement grid. Each individual element shows a distribution from approximately 10 - 35 at.% over the substrate, i.e. there are five main composition gradients. However, these 342 MAs cover only 0.5 % of the complete composition range (considering all combinations in 5 at.% steps) and if the search space is restricted to the center region of a quinary system (10 - 35 at.% variation for each element, "HEA region"), i.e. there is no dominating element, the coverage for one ML is about 9%. This is on the one hand a large composition space compared to what can be achieved by single sample synthesis, on the other hand it is small compared to the complete composition range of a single quinary system.

Therefore, we propose to explore this HEA region much more efficiently by a deposition source permutation strategy. In principle, two strategies are possible for extending the covered composition space, i.e. surpassing the compositions covered by one ML. A first option is to create further MLs by variation of the deposition rate of individual deposition sources. In this way, the multidimensional composition space is sliced with different contributions of individual elements (i.e. different amounts of deposited material from individual elements). However, for a fixed target arrangement, the elements of the neighboring (or opposed) deposition sources are correlated as well in the resulting compositional gradients, which restrict the complete coverage of the system. For example, there are no positions on the ML for which the concentration of a single element can be changed independently, without changing the relative composition of the remaining four elements on the co-sputtered ML. Furthermore, the composition-property trends observed on a single ML can lead to false assumptions about the underlying compositional trends due to these correlated composition gradients.

A second and new option proposed here is to synthesize MLs using a sequence of permutations of the arrangement of the deposition sources, for which the relative alignment of composition gradients is changed for each ML deposition, i.e., the multidimensional composition space is sliced at different angles for each permutation. Figures SI 10-15 show pairplots of the relative compositions of all binary combinations in a quinary system for two MLs that were synthesized using different deposition source arrangements, showing the presence and absence of certain combinations of compositions in a single ML. In case of five sources, 24 target arrangements are possible, keeping one source at a fixed position (P(n) = 4!). To minimize the number of necessary experiments to cover a substantial amount of a quinary composition space, we use the distribution of each element over the MAs of an initial ML and simulate the compositions which are covered by 23 following deposition source permutations. We rank all permutations against the initial ML by calculating the number of obtained *additional* unique compositions. Thus, the permutation with the least overlap with the previous ML is the next to be fabricated. Following this scheme, all remaining permutations are ranked iteratively against all previous permutations. Based on this ranking, we guide the synthesis of additional MLs (see cathode arrangements in figure SI2). The composition space of a quinary system can be visualized in 2D or 3D projections of the regular 5-cell (pentachoron, 4-simplex) by using generalized barycentric coordinates. Figure 1C shows the barycentric plot of two permutated quinary MLs on a 3D projection of a regular 5-cell. Although this is a purely illustrative visualization, it conveys a sense for the scale of a single ML with respect to the five-dimensional composition space. Following the permutation of deposition source arrangements different slices of the 5D composition space are efficiently produced.

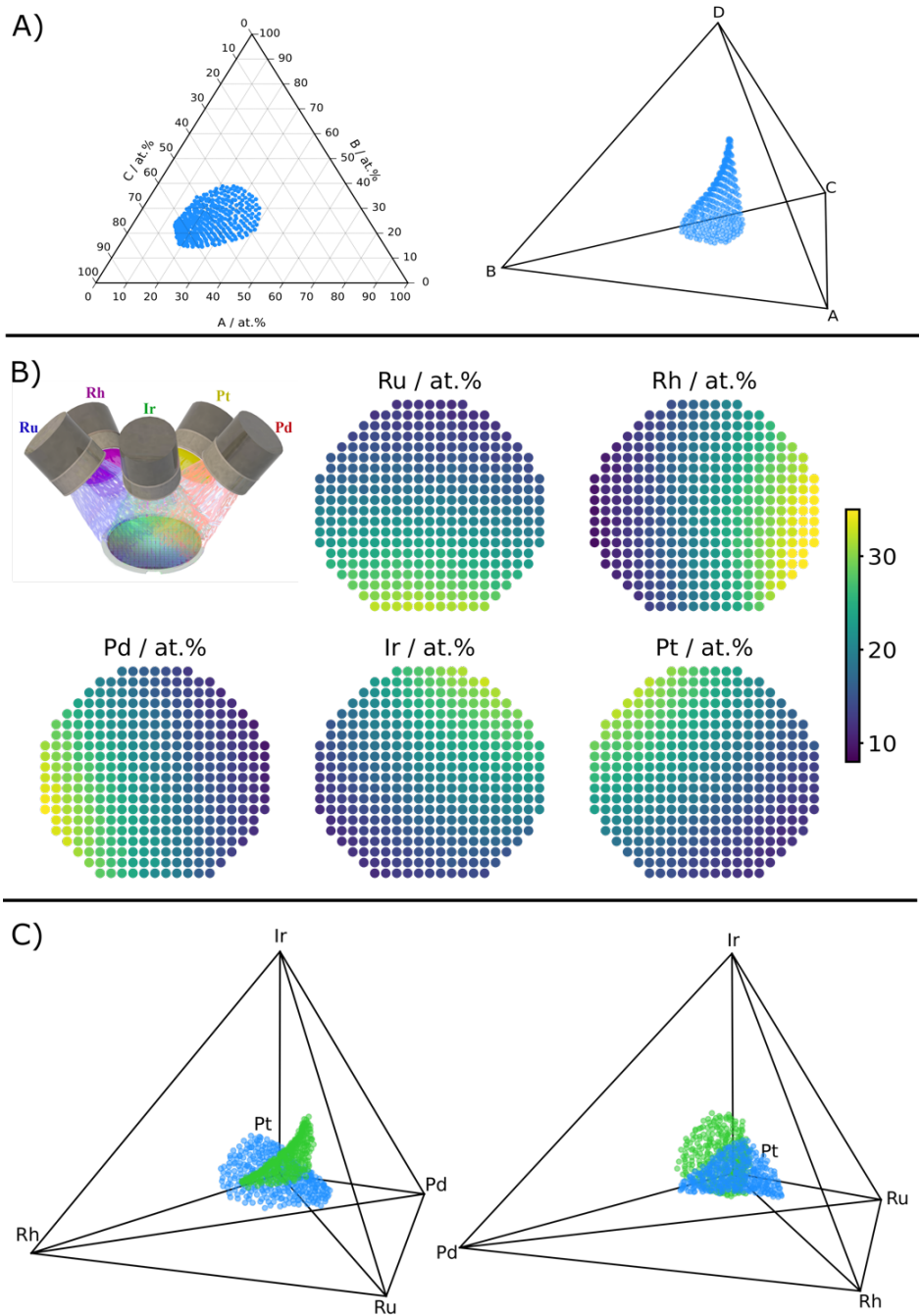

*Figure 1: Visualization of the compositional coverage of continuous composition spread materials libraries co-sputtered from three to five sources forming ternary to quinary systems. In all cases, 342 measurement areas (on a regular grid 4.5 mm by 4.5 mm over a 100 mm diameter substrate are visualized: A) Coverage of a ternary and a quaternary materials library. B) Target arrangement and compositional gradients of a co-sputtered quinary materials library. C) Visualization of the five-dimensional composition space of two co-deposited quinary materials libraries in a 3D projection of a regular 5-cell (blue: initial target arrangement, green: permutated target arrangement; left/right: different rotations of the plot).*

Figure 2a and b show the simulated coverage (cumulated compositions) of the full composition space and the HEA composition space in dependence of the number of deposited MLs, each with a different permutation, i.e. target arrangement. These ML depositions become ineffective after ten permutated depositions, as the increase in covered composition space becomes smaller for each successive permutation, i.e. the information gain of the second permutation is comparable to that of permutations 10-24 combined. Depositions performed for all possible permutations would result in a coverage of 3.2 % of all possible unique compositions in the full and up to approximately 50% in the HEA composition space, considering 5 at.% steps. In order to limit experimental efforts, we choose to deposit ML for the first six permutations covering approximately 30 % of the HEA composition space (10 to 35 at.%) (see deposition source arrangements in figure SI2 and composition maps in figures SI3-9). Figure 2c and d illustrate the covered compositions (red points) respective to the compositions in the reduced quinary space (grey points).

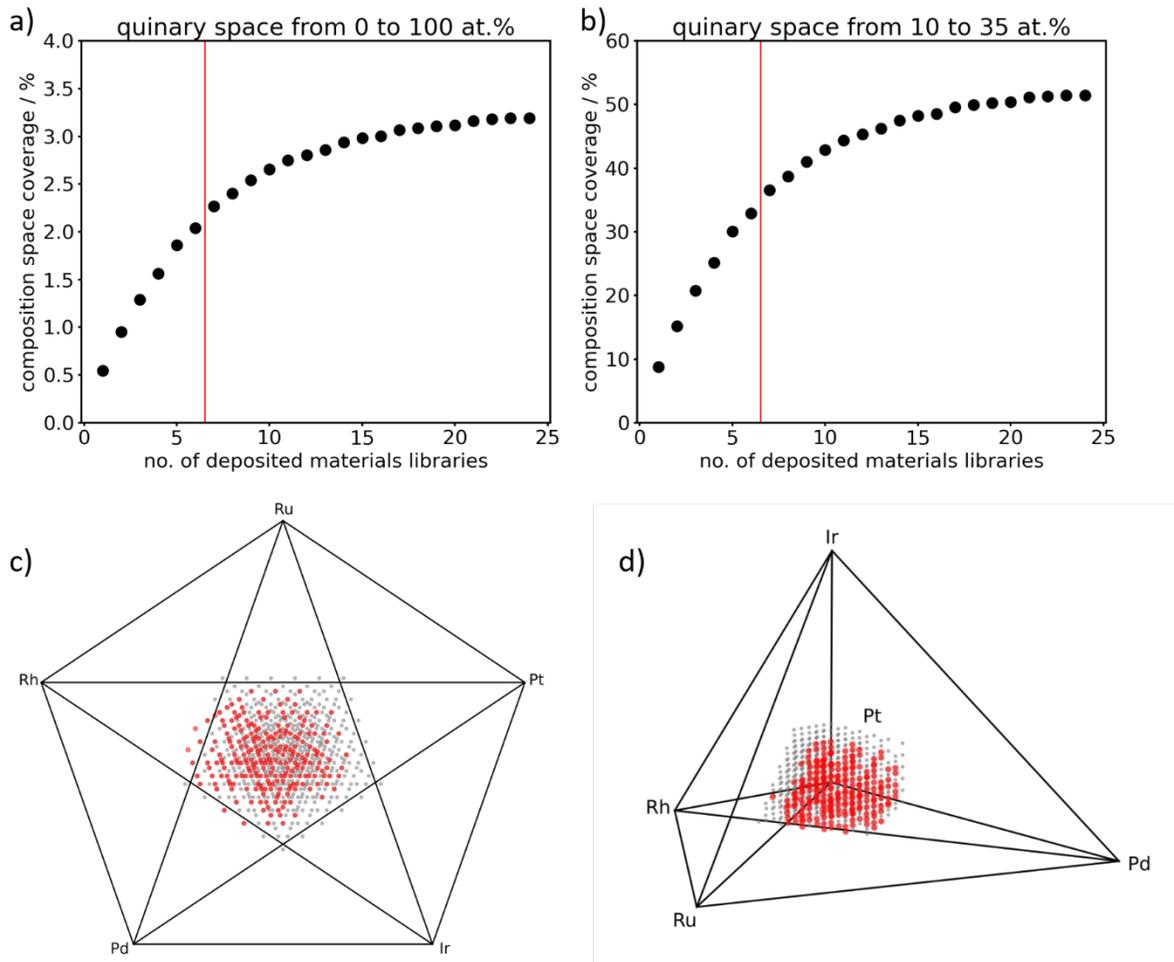

*Figure 2: Composition space coverage of quinary materials libraries deposited with all 24 possible deposition source arrangements. a) Number of deposited materials libraries and composition space coverage of the full quinary composition space sampled in 5 at.% steps. b) Composition space coverage of the reduced HEA composition space (10 to 35 at.%). c, d) Visualization of the covered compositions from six materials libraries that were deposited using six permutated deposition source arrangements in the HEA composition space in a 2D(c) and 3D(d) projection of a regular 5-cell. Grey points denote all possible compositions of the reduced HEA composition space and the red points the compositions measured in the synthesized materials libraries.*

Analysis of the ORR activity using supervised and unsupervised machine learning

The six permutated MLs were screened for the oxygen reduction reaction (ORR) in acidic environment (0.1 M $HClO_4$) using a high-throughput scanning droplet cell (SDC). This setup preserves the same measurement conditions over hundreds of MAs, thus allowing a credible comparison of electrochemical activities between different MAs and MLs. Linear sweep voltammograms were recorded on all MAs. Figure SI16 shows electrocatalytic activity maps obtained by plotting current densities recorded at a potential of 0.8 V vs. reversible hydrogen electrode (RHE) at the positions of the MAs on the ML. While trends in electrochemical activity are evident on MLs 1-4, ML5 and ML6 do not show significant trends in activity at the overpotential of 422.9 mV vs the formal potential of the ORR (i.e. 800 mV vs. RHE). However, a clear maximum activity is observed on ML4. Analysis of the total dataset shows that the highest activities are achieved for compositions that are high in Ru and Pd and show lower contributions of the remaining elements. The composition with the highest activity is $Ru_{25}Rh_{15}Pd_{31}Ir_{15}Pt_{14}$. Spearman correlation scores were calculated for the complete dataset, revealing that indeed Ru and Pd show a positive correlation with the ORR activity and Rh shows the highest negative score (Figure SI19). An additional ML of the same target arrangement as ML4 but decreased Rh content (ML4b) shows an increased activity for the ORR compared to ML4. The highest activity is observed for $Ru_{17}Rh_5Pd_{19}Ir_{29}Pt_{30}$ (Figure SI23).

A linear regression model was fitted to the total dataset (excluding ML4b) to predict the electrochemical activity for the complete quinary composition space. The highest activity for the total composition space is predicted for binary Ru-Pd compositions. An independent, combined computational and experimental study identified a local minimum in the ORR activity for $Ru_{35}Pd_{65}$, which was validated experimentally[27]. In direct comparison between the experimental Ru-Pd composition spread and the most active composition identified on ML4b, however, the quinary composition exceeds the highest predicted activity of both DFT model and experimental Ru-Pd.

To investigate the materials parameters causing the disagreement between both models and the experimentally observed ORR activity, we characterize each ML using high-throughput X-ray diffraction (XRD) and perform non-negative matrix factorization (NMF) on 2052 XRD patterns to break the dataset down to an interpretable representation. NMF identifies four representative patterns that are weighed for each experimental XRD pattern (Figure 3A). Kernel principal component analysis is applied on the NMF weights and the chemical composition of the dataset for the visualization of two-dimensional composition-structure-property maps (Figure 3). Based on the inspection of the NMF representative XRD patterns and their respective weights, a region of predominant nanocrystalline single phase fcc solid solution was identified (compare phase mapping results shown in Figure SI25). Within this region, certain chemical compositions (high Ir and Pt) cause a local maximum in electrochemical activity for ORR (see Figure 3B and C for an applied potential of 800 mV and 600 mV vs. RHE, respectively). An approximately 10-fold ORR current increase is observed for $Ru_{17}Rh_5Pd_{19}Ir_{29}Pt_{30}$ compared to the lower activity compositions dominated by Pd and Rh.

As both LR and computation model do not consider structural factors, the validity of the models is restricted to compositional subspaces for which the structural factors remain unchanged. While in the LR model fitted to the experimental data structural factors are indirectly encountered for in the measured electrochemical activity, the computational model assumes an ideal structure (fcc solid solution, (200)-surface). In order to combine the advantages of both, computational model and machine learning on experimental data, a future strategy could be to evaluate the computational model previous to the experiment to synthesize different target permutations centered around the computed, local maxima. From this experimental dataset, machine learning can be applied to iteratively optimize chemical compositions, considering both, structural and chemical factors of combinatorial datasets.

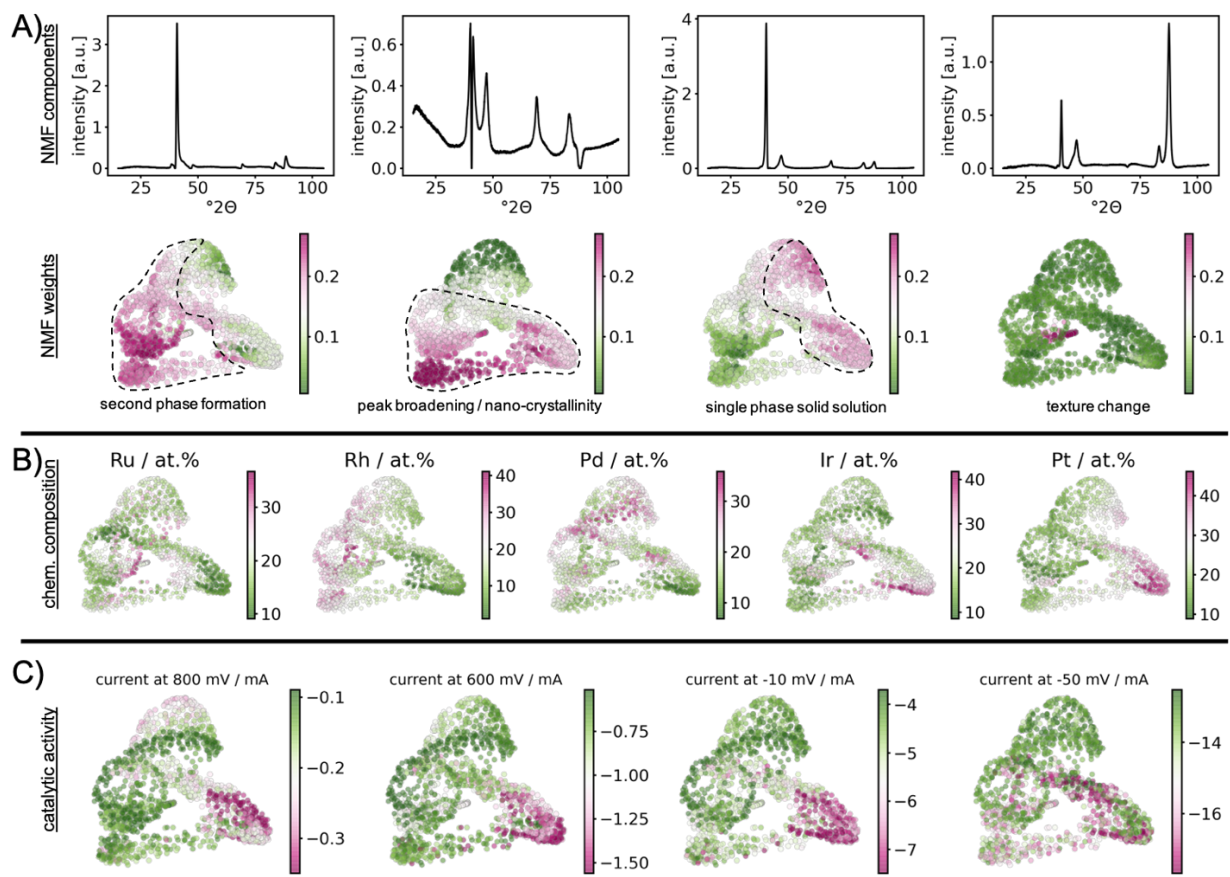

*Figure 3: Results of unsupervised machine learning visualized as composition-structure-property maps. A: Representative NMF patterns obtained from the XRD dataset and their respective weights over the kernel-PCA composition-structure-representation. B: composition-structure-representation color-coded by chemical composition. C: electrochemical current for 800, 600, -10 and -50 mV vs. RHE color-coded over the composition-structure-representation.*

In summary, we demonstrate a strategy to effectively cover substantial parts of a quinary composition space with reasonable experimental efforts using a permutation strategy for the synthesis of composition spread materials libraries. High-throughput characterization was applied to map chemical, structural and electrochemical properties. The generated dataset contains viable information on the influence of chemical composition on the electrochemical activity that allow for iterative optimization of the electrochemical activity for the oxygen reduction reaction. Manifold analysis of the dataset reveals that both, structural properties and chemical composition determine electrochemical activity. A future strategy is proposed to leverage the advantages of computational models and machine learning on experimental datasets.

## Methods

Synthesis of materials libraries (MLs)

A total of seven quinary Ru-Rh-Pd-Ir-Pt MLs (M1–M6, ML4b) were co-sputtered from a loadlock-equipped combinatorial magnetron sputtering system (DCA Instruments, Finland). Except for target position changes which led to six different permutations, all other experimental parameters used during sputtering were the same. In a typical ML synthesis, high purity (Rh: 99.95%, Ir: 99.9%, Pd: 99.99%, Pt: 99.99%, Ru: 99.95%) 100 mm diameter Ir, Pd, Pt and Ru targets were used. For cost reasons, the Rh target was 51 mm diameter on a same-sized cathode. Each target was positioned with an inclination angle of around 45° with respect to the substrate. The deposition rate was adjusted for each target in order to get equiatomic composition in the center of ML. The deposition was carried out without intentional heating. The cathode tilts of Rh, Ru and Pd were set to 90° during the deposition of ML4b. 100 mm diameter sapphire wafers (c-plane) were used as substrates. Prior to each deposition, the chamber vacuum was on the order of $10^{-5}$ Pa. The targets were pre-sputtered for 30 s to clean their surface. During deposition, the chamber pressure was set to 0.667 Pa (Ar: 99.9999%), and the substrate was kept stationary for 5 mins to obtain five continuous compositional gradients. After deposition, the film thickness was determined to be around 100 nm.

Electrochemical characterization

All MLs were analyzed using a scanning droplet cell (SDC), which allows localized characterization. All electrochemical measurements were conducted in 0.1 M $HClO_4$ electrolyte in a three-electrode system containing an Ag|AgCl|3M KCl and Pt wire as a reference and counter electrode, respectively. The working electrode is formed in every spot where the tip of the SDC touches the sample surface. The tip opening size (7.35 x $10^{-3}$ $cm^{-2}$) determines the working electrode surface area. Linear sweep voltammetry was performed between ca. 1 V and 200 mV vs. reversible hydrogen electrode (RHE) with a scan rate of 10 mV $s^{-1}$. All potentials are reported versus the RHE according to the following equation:

$U_{RHE}$ (V)= $U_{(Ag|AgCl|3M\ KCl)}$ + 0.210 + (0.059·pH),

where $U_{(Ag|AgCl|3\ M\ KCl)}$ is the potential measured vs. Ag|AgCl|3M KCl reference electrode, 0.210 V is the standard potential of the Ag|AgCl|3M KCl reference electrode at 25 °C. 0.059 is the result of $(RT)·(nF)^{-1}$, where R is the gas constant, T is the temperature (298 K), F is the Faraday constant and n is the number of electrons transferred during the reaction. Particular MAs on MLs are separated from each other by 4.5 mm (342-MA grid per ML).

Energy dispersive x-ray spectroscopy

Automated energy dispersive X-ray spectroscopy (EDX) was performed at 20 kV acceleration voltage in a scanning electron microscope (JEOL 5800) using an INCA X-act detector (Oxford Instruments). 342 measurement areas on a 4.5 mm x 4.5 mm coordinate grid were measured. The magnification for each EDX-scan was 600x.

X-ray Diffraction

High-throughput X-ray diffraction (XRD) was made from a Bruker D8 Discover (Bruker Corporation), Bragg−Brentano geometry, equipped with a VANTEC-500 area detector, Cu Kα radiation, sample to detector distance = 149 mm. The X-ray beam size was collimated to 1 mm in our experiment. (collimator

diameter = 1 mm with a divergence of below 0.007°). An area detector was used to collect the diffraction data, and three frames were collected for each measured area. These 3 frames were integrated into one-dimensional diffractograms using the DIFFRAC.EVA software (Bruker Corporation) and were then used for phase identification. The results of the phase identification are shown in figure SI25.

Data analysis and machine learning

Data analysis was performed in Python 3.7.9. Scikit-learn version 0.23.2 was used for linear regression, non-negative matrix factorization, mutual information regression and kernel principle component analysis. Spearman correlation analysis was done using the pandas package version 1.2.1.

**Data availability**

The data used in this study is available from the responding author on reasonable request.

**Code availability**

The code for data analysis and visualization is available from the corresponding author on reasonable request.

**Contributions**

LB conceptualized the study and performed data analysis and machine learning. OK performed the electrochemical characterization and advised the data analysis. LB, OK and AL wrote the main parts of the manuscript. BX, AS and LB synthesized the material libraries. BX performed EDX and XRD measurements. WS and TL advised the electrochemical characterization and data analysis. TL advised the visualization concept. JP and JR performed electrochemical calculations. All authors contributed to writing the manuscript.

**Acknowledgements**

A.L. and L.B. acknowledge funding by the German Research Foundation (DFG) as part of Collaborative Research Centers SFB-TR 87. W.S. acknowledges funding from Deutsche Forschungsgemeinschaft (DFG) under Germany's Excellence Strategy (EXC 2033-390677874–RESOLV). We thank Sabrina Baha for assistance with accompanying measurements. A.L., A.S., and B.X. acknowledge funding from DFG projects LU1175/22-1 and LU1175/26-1. The center for interface dominated high-performance materials (ZGH, Ruhr-Universität Bochum, Bochum, Germany) is acknowledged for X-ray diffraction experiments.

# Supporting Information

## Combinatorial materials discovery strategy for high entropy alloy electrocatalysts using deposition source permutations

Lars Banko[1,#], Olga A. Krysiak[2,#], Bin Xiao[1], Tobias Löffler[1,2,3], Alan Savan[1], Jack Kirk Pedersen[4], Jan Rossmeisl[4], Wolfgang Schuhmann[2,], Alfred Ludwig[1,3,*]

5. Materials Discovery and Interfaces, Institute for Materials, Ruhr-University Bochum, 44801 Bochum, Germany
6. Analytical Chemistry – Center for Electrochemical Sciences (CES), Faculty of Chemistry and Biochemistry, Ruhr University Bochum, Universitätsstr. 150, D-44780 Bochum, Germany
7. Center for Interface-Dominated High Performance Materials, Ruhr-University Bochum, 44801 Bochum, Germany
8. Theoretical Catalysis – Center for High Entropy Alloy Catalysis (CHEAC), Department of Chemistry, University of Copenhagen, Universitetsparken 5, 2100 Copenhagen, Kbh (Denmark)

[#] equal contributions

[*] corresponding author

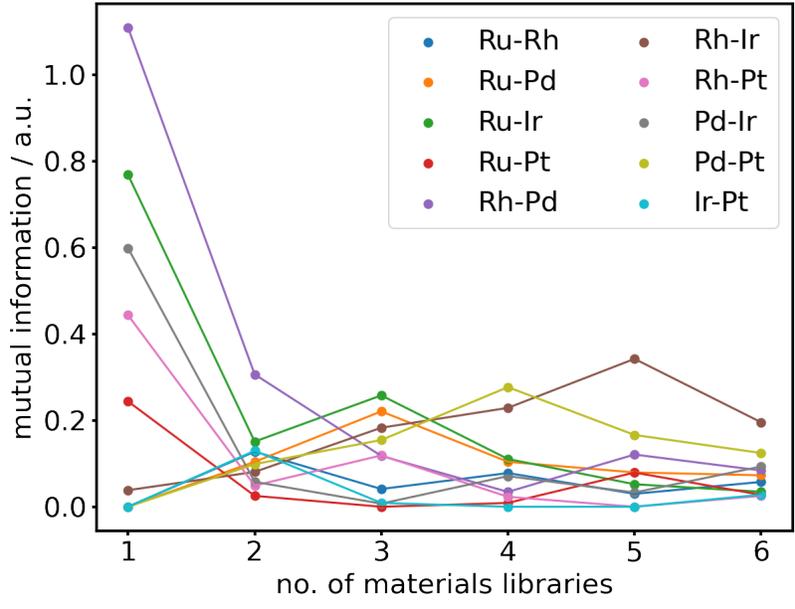

*SI 1: Results of mutual information regression. Values are zero if two random variables are independent of each other and high values indicate dependence. The results show that the mutual information is high for the first permutation for elements that were deposited from opposite deposition sources. The mutual information score decreases with further permutations.*

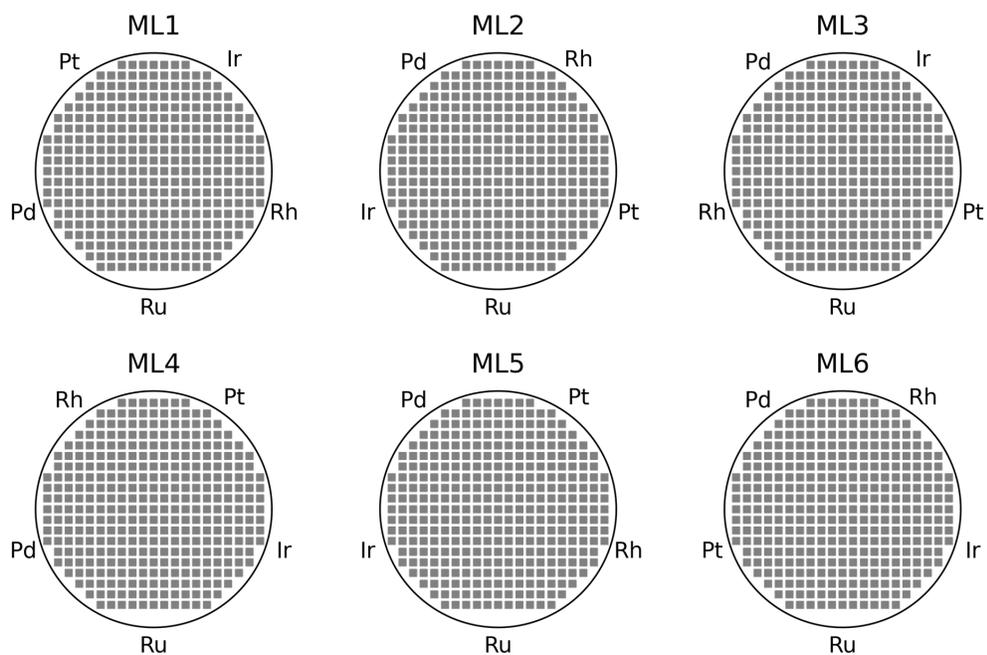

SI 2: Deposition source / target arrangements of the deposited materials libraries. The measurement coordinates of the substrate are plotted (grey squares) and the position of the deposition source relative to the substrate are indicated.

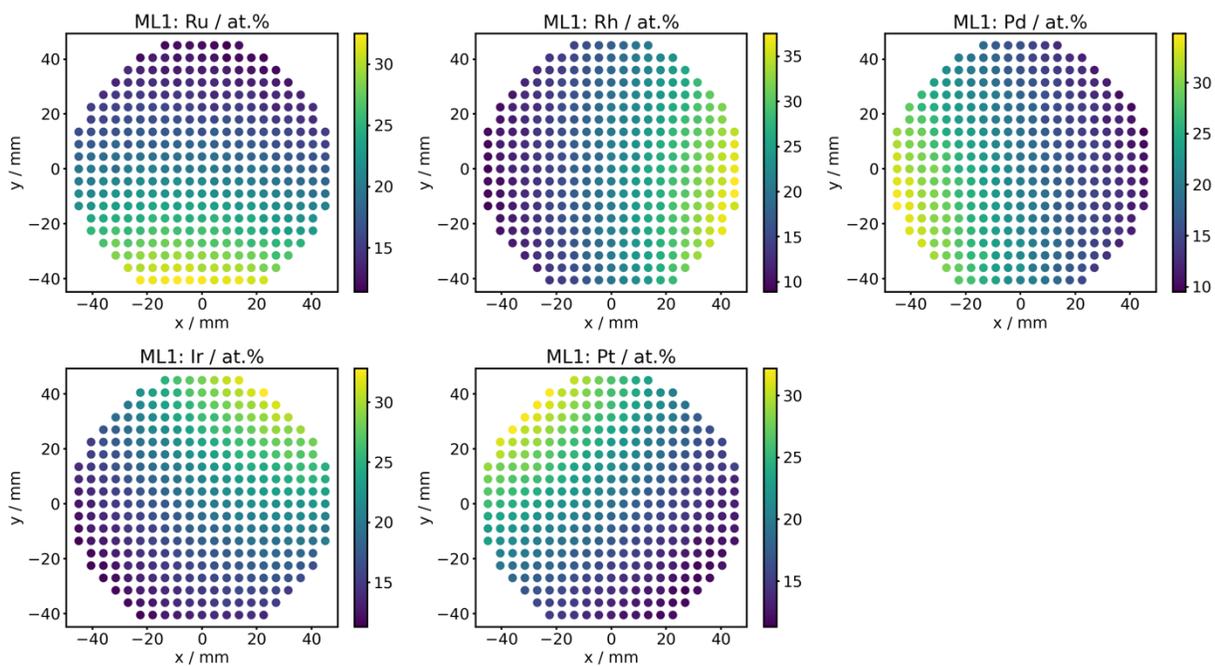

SI 3: Chemical composition as measured by EDX color-coded over the substrate coordinates for ML1.

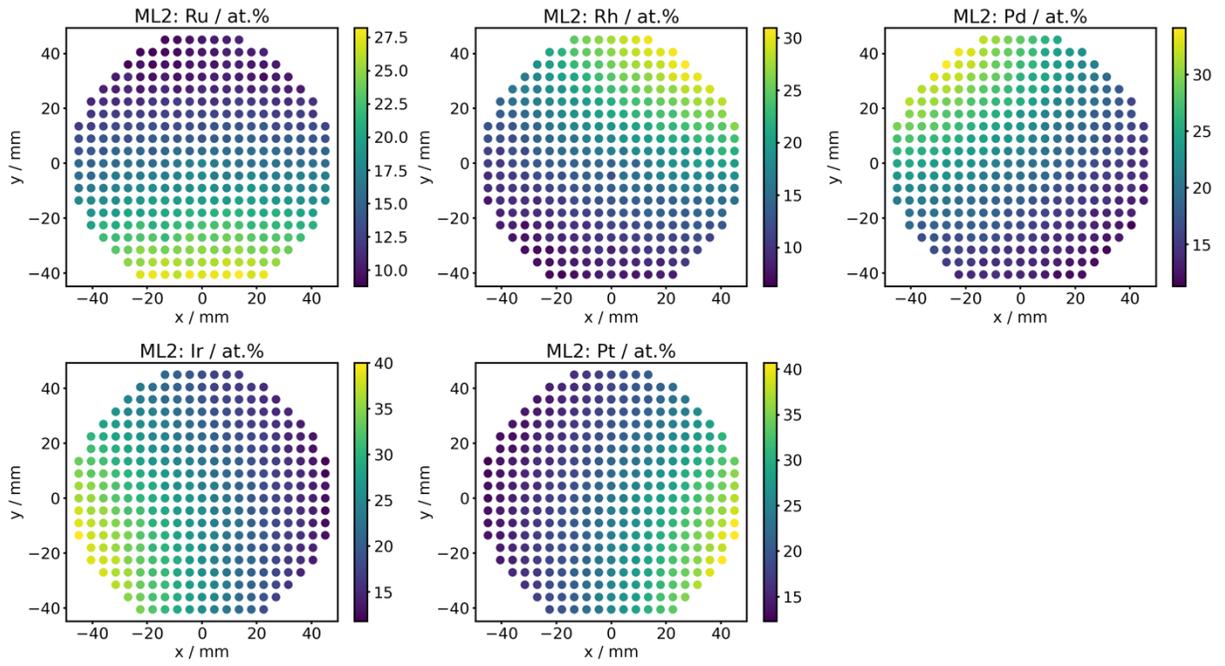

*SI 4: Chemical composition as measured by EDX color-coded over the substrate coordinates for ML2.*

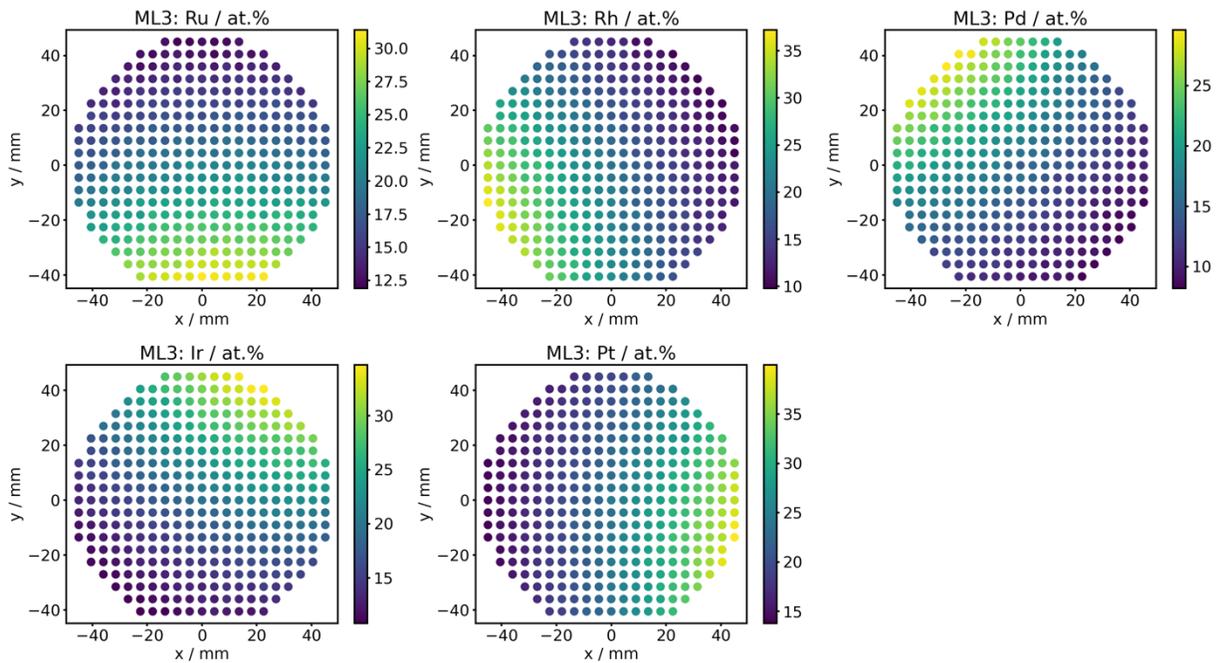

*SI 5: Chemical composition as measured by EDX color-coded over the substrate coordinates for ML3.*

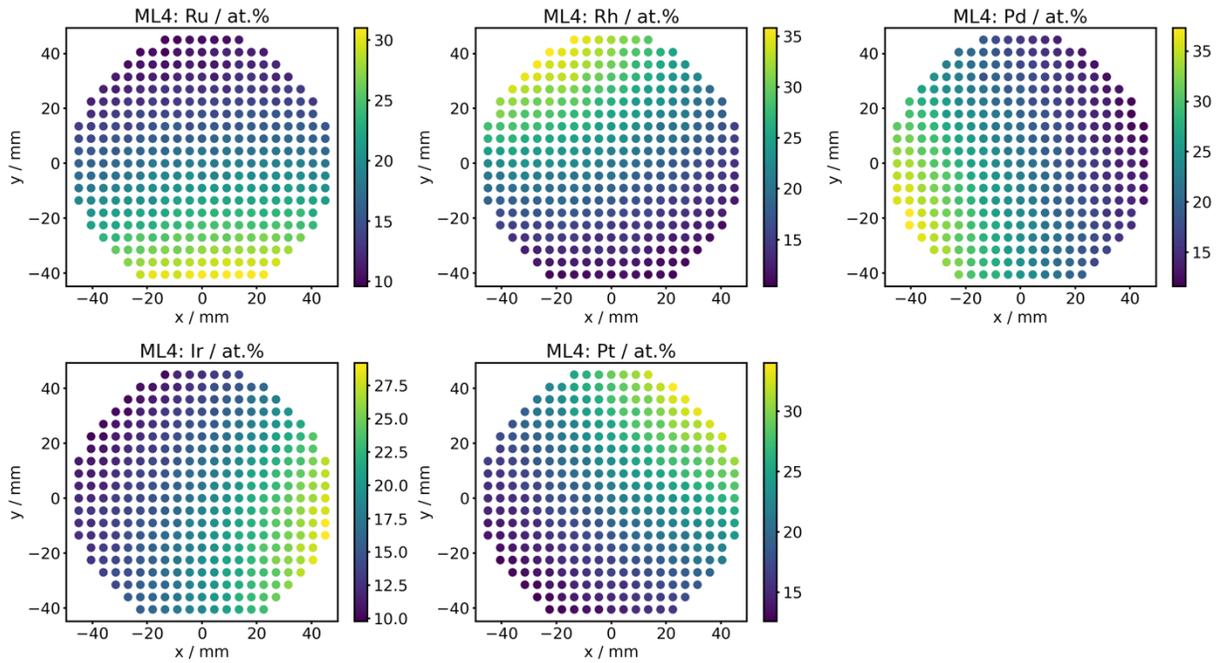

*SI 6: Chemical composition as measured by EDX color-coded over the substrate coordinates for ML4.*

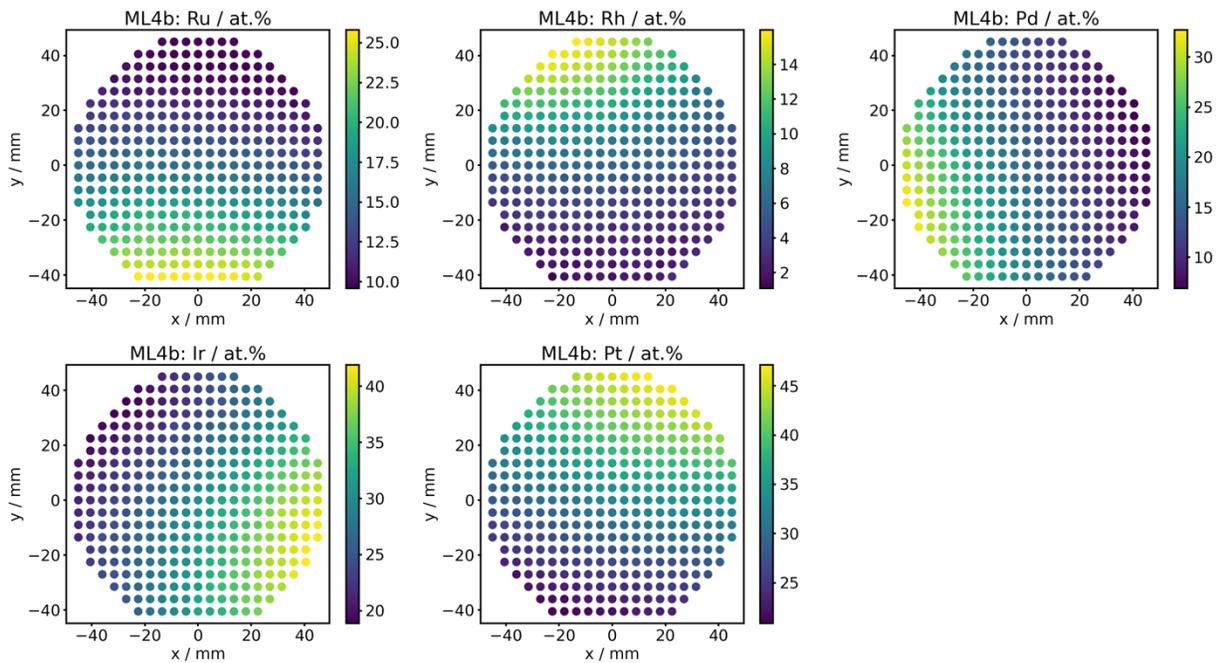

*SI 7: Chemical composition as measured by EDX color-coded over the substrate coordinates for ML4b.*

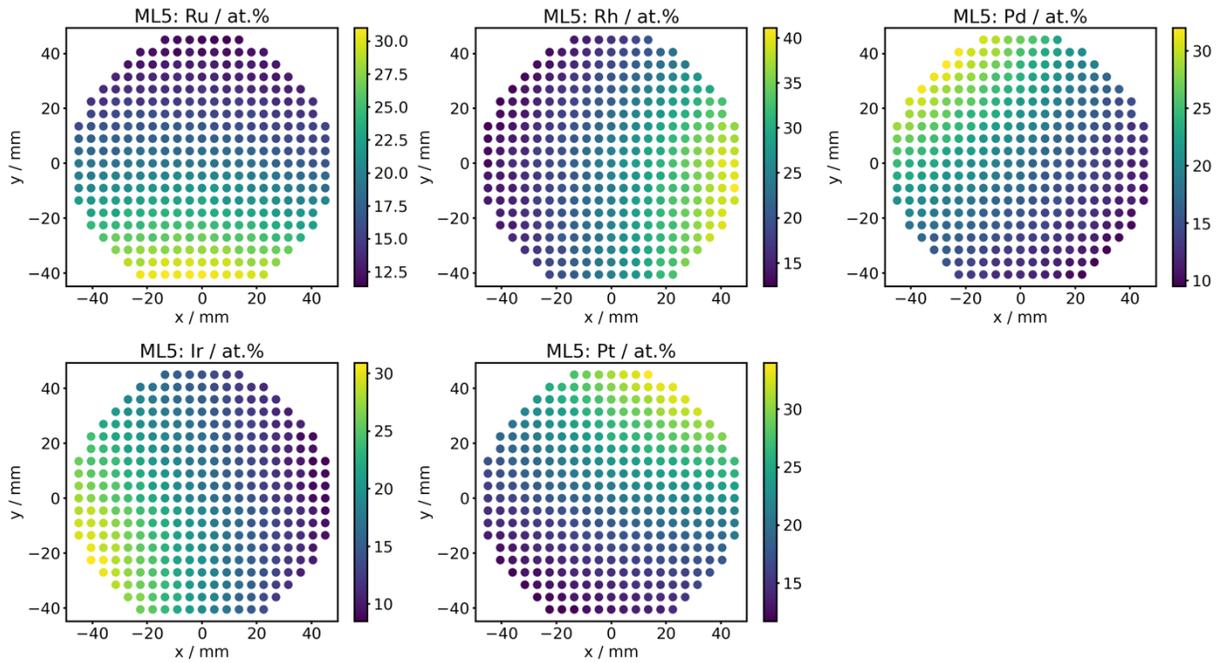

*SI 8: Chemical composition as measured by EDX color-coded over the substrate coordinates for ML5.*

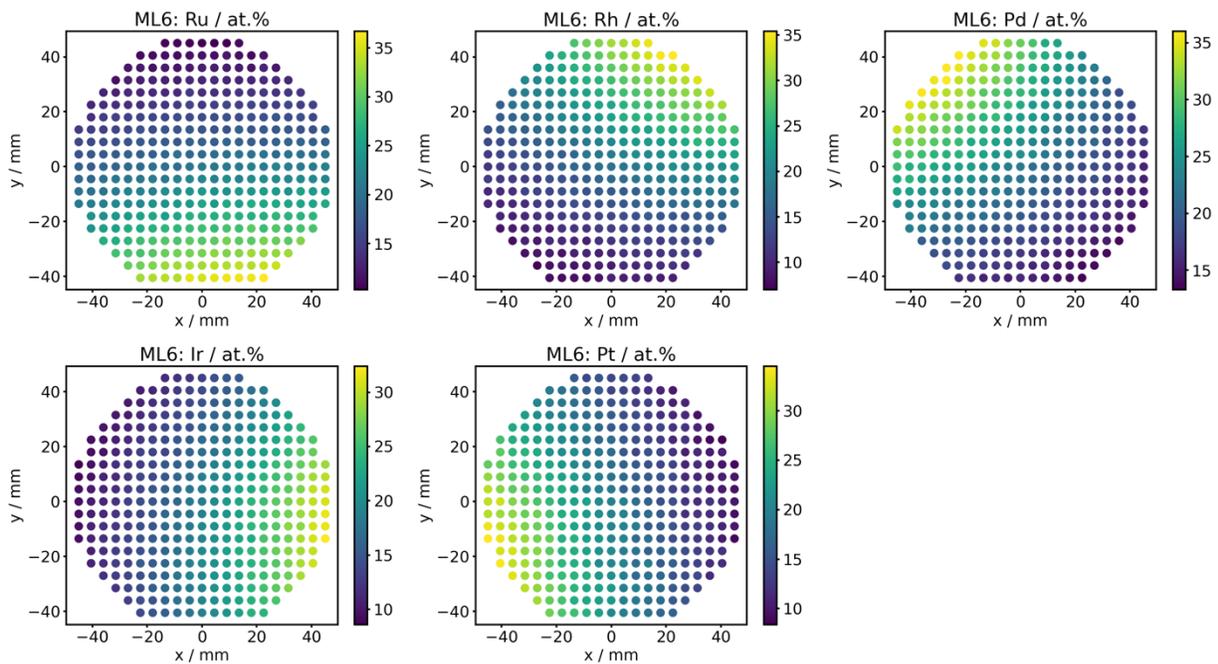

*SI 9: Chemical composition as measured by EDX color-coded over the substrate coordinates for ML6.*

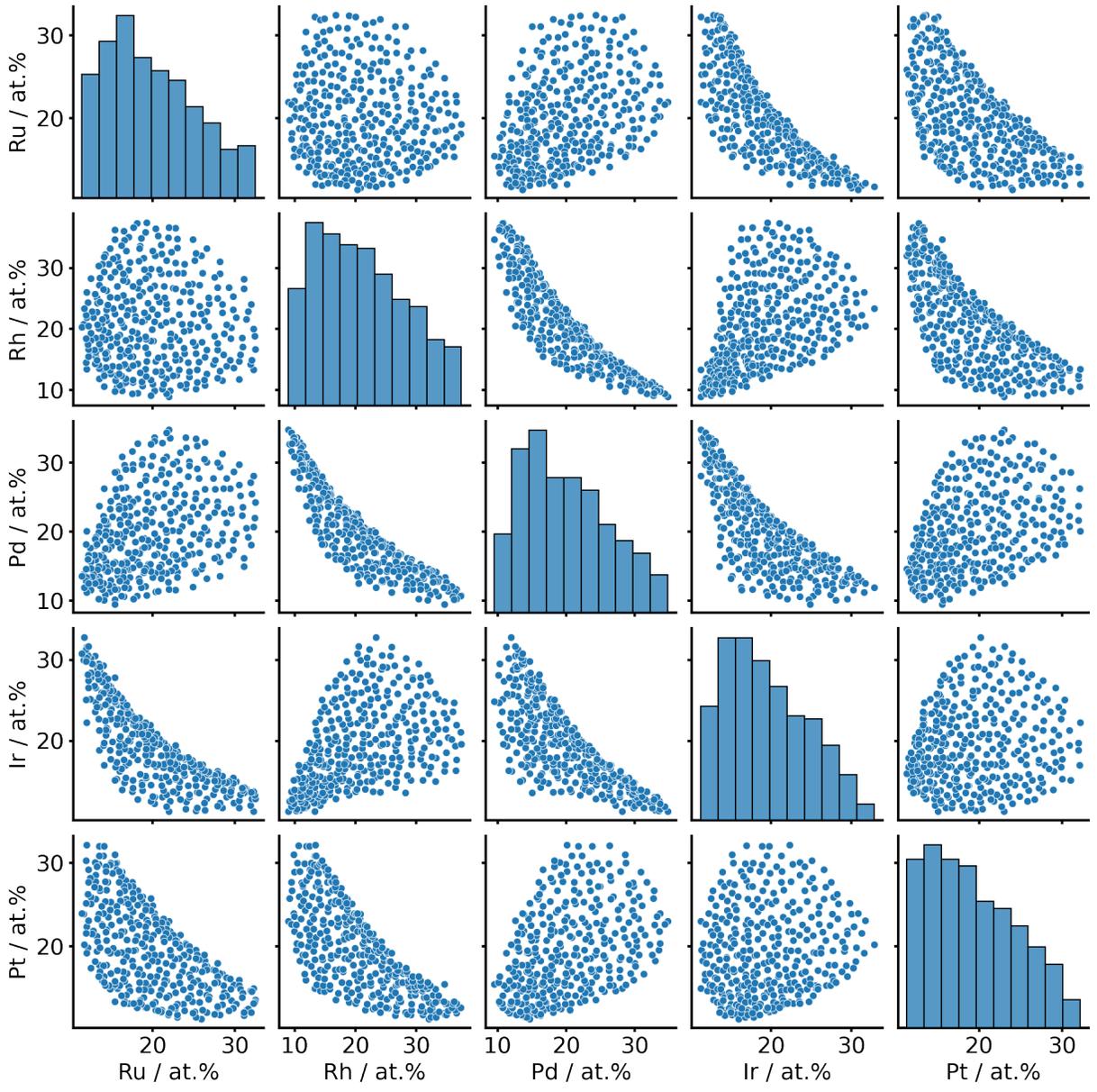

SI 10: Pairplot showing all binary combinations of ML1.

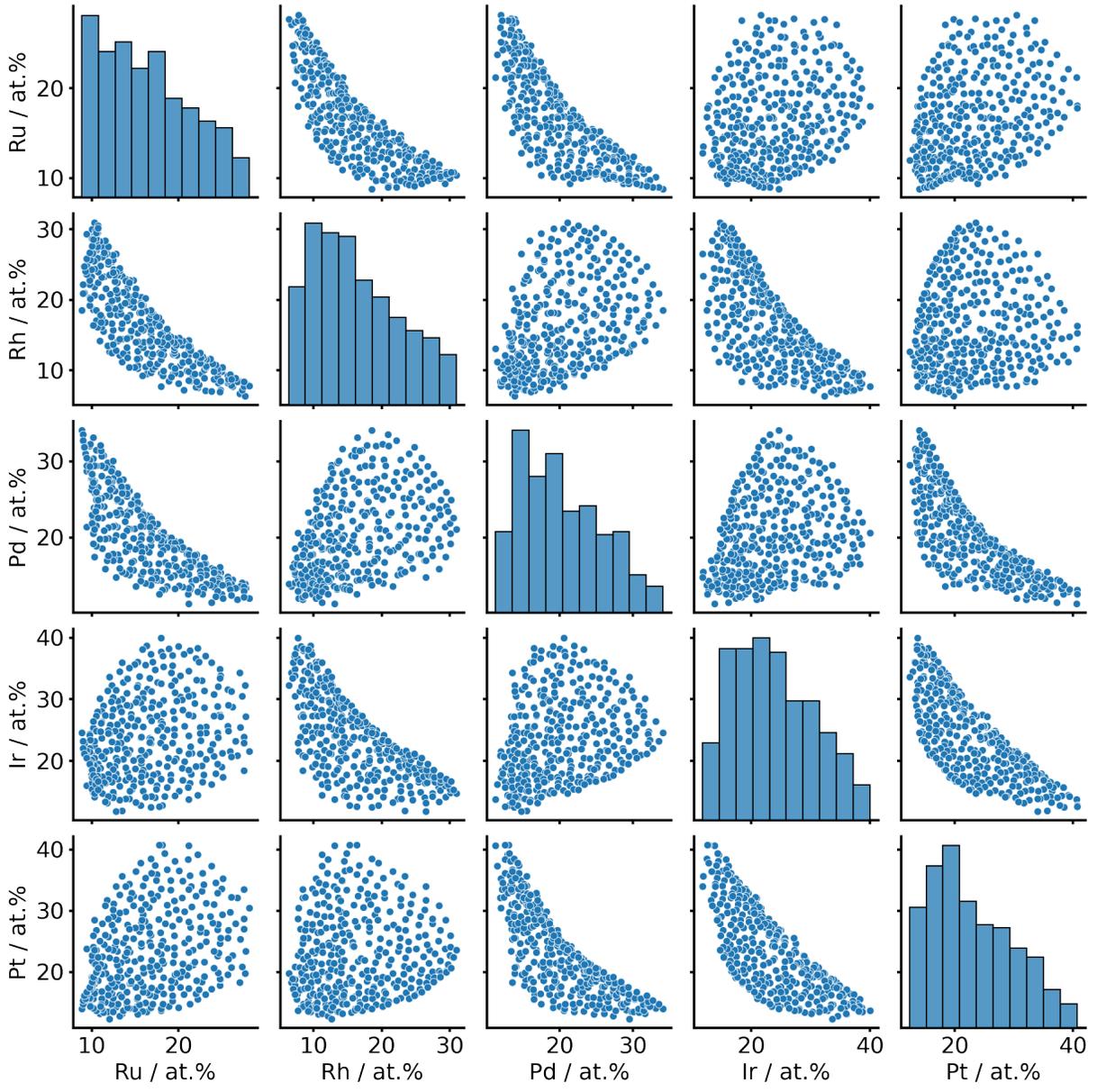

*SI 11: Pairplot showing all binary combinations of ML2.*

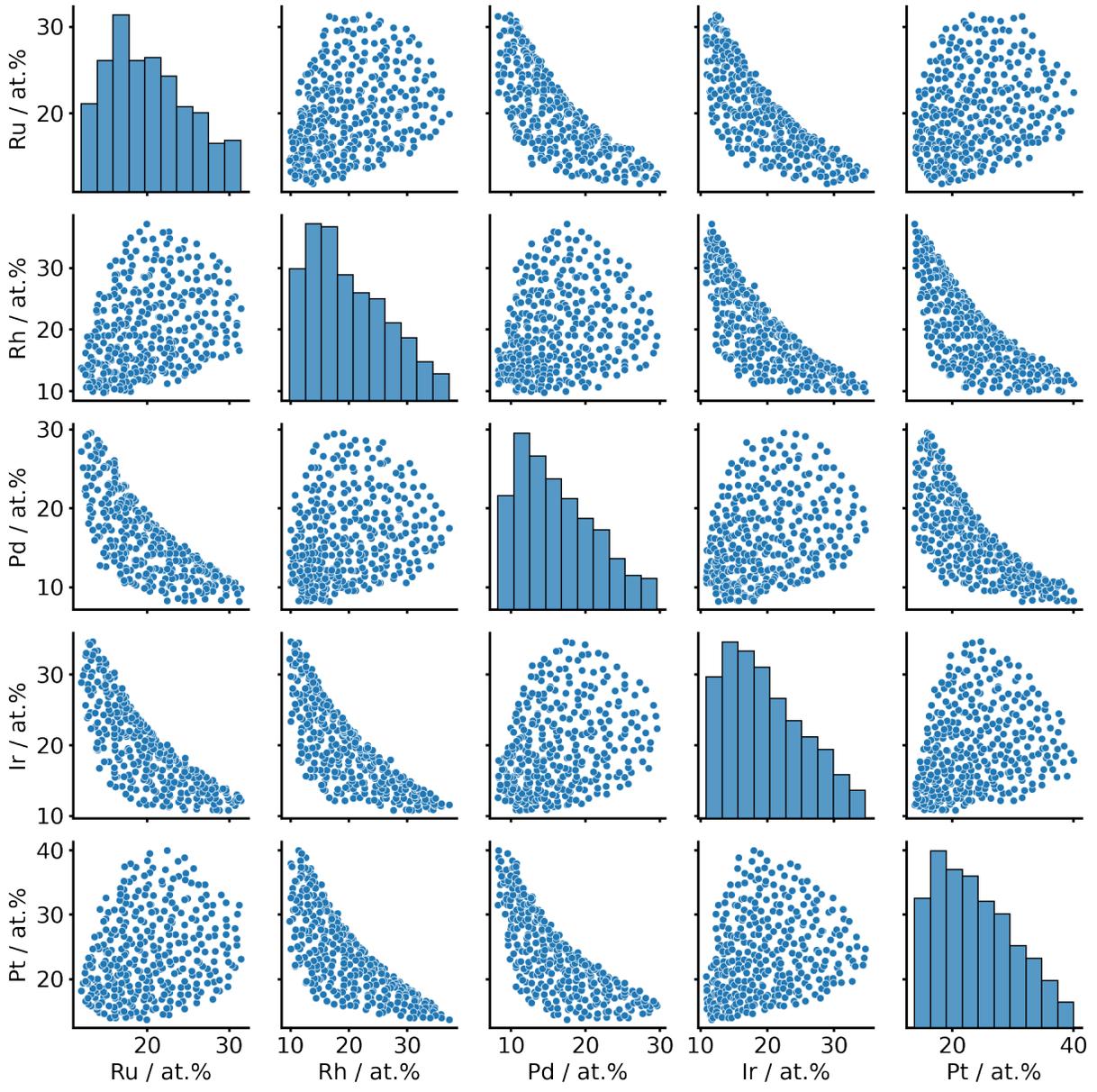

SI 12: Pairplot showing all binary combinations of ML3.

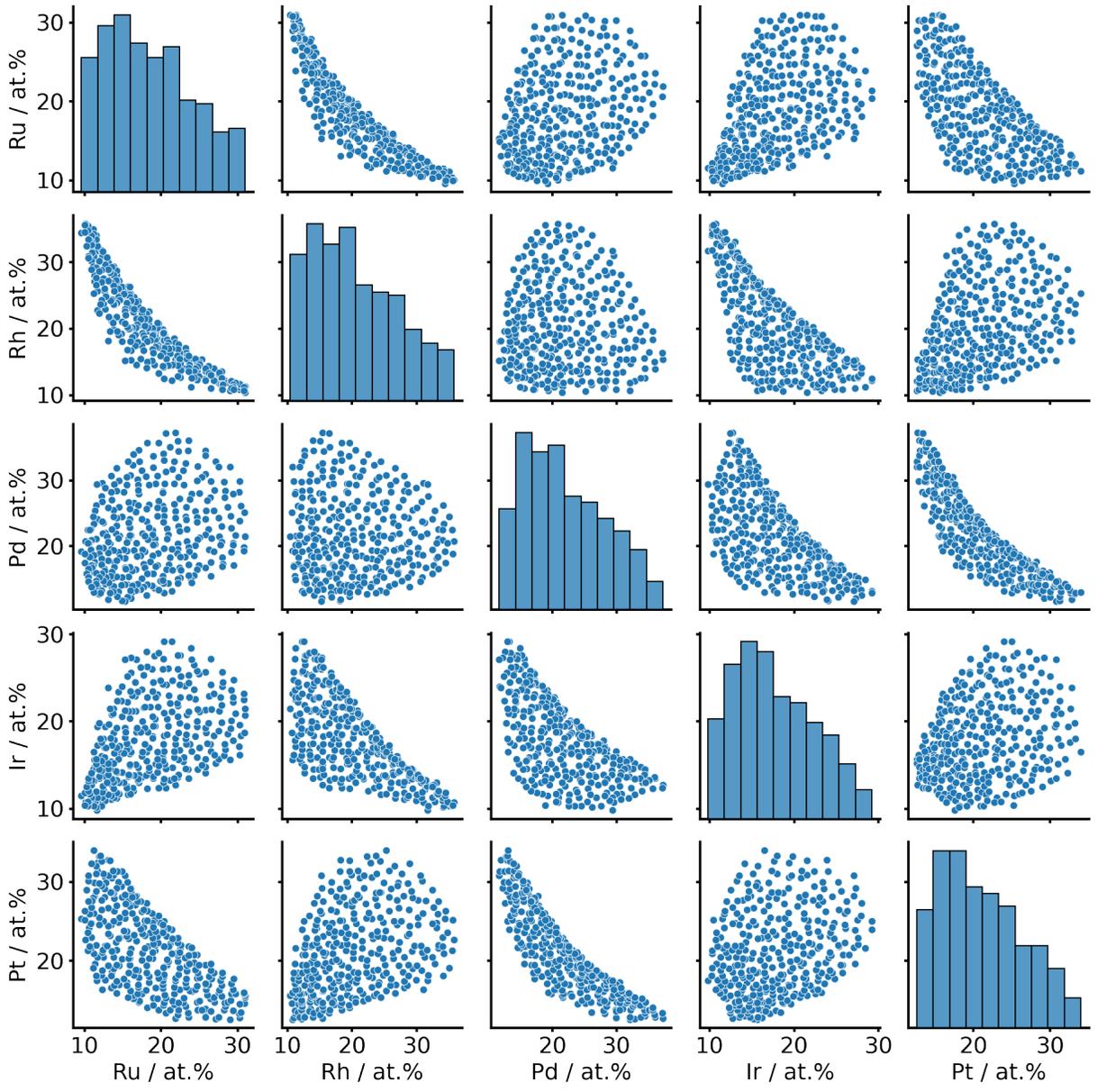

SI 13: Pairplot showing all binary combinations of ML4.

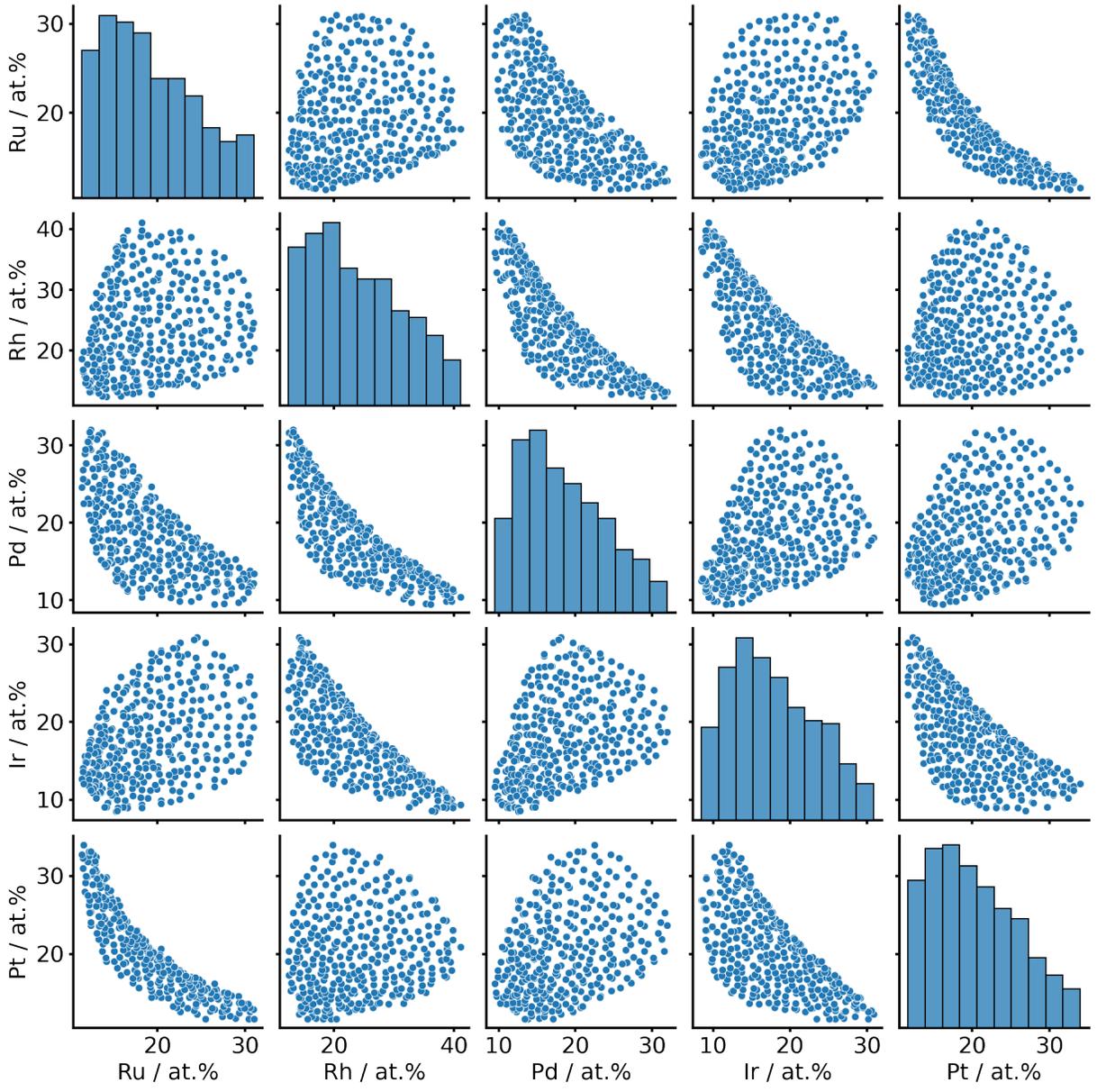

SI 14: Pairplot showing all binary combinations of ML5.

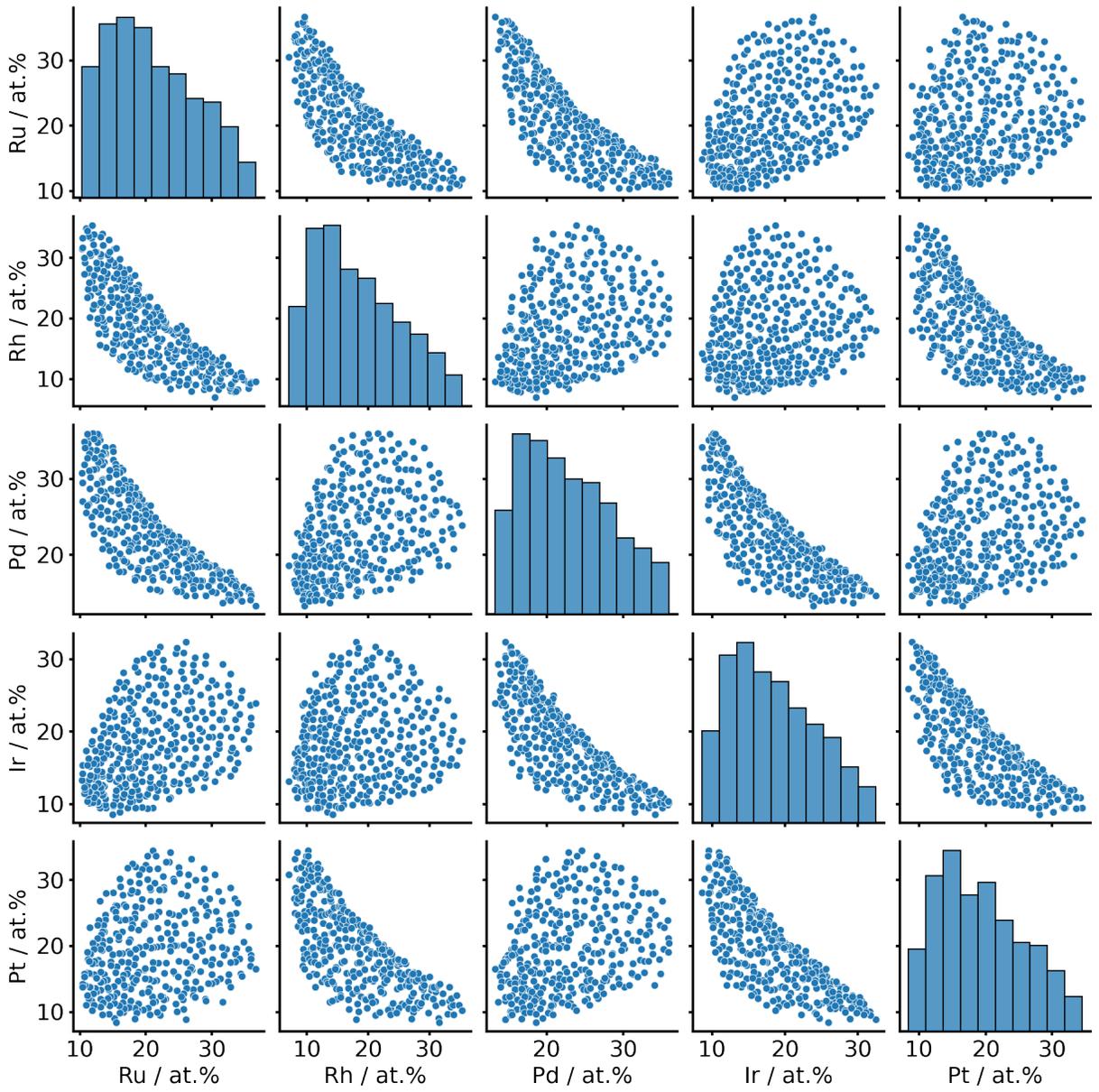

SI 15: Pairplot showing all binary combinations of ML6.

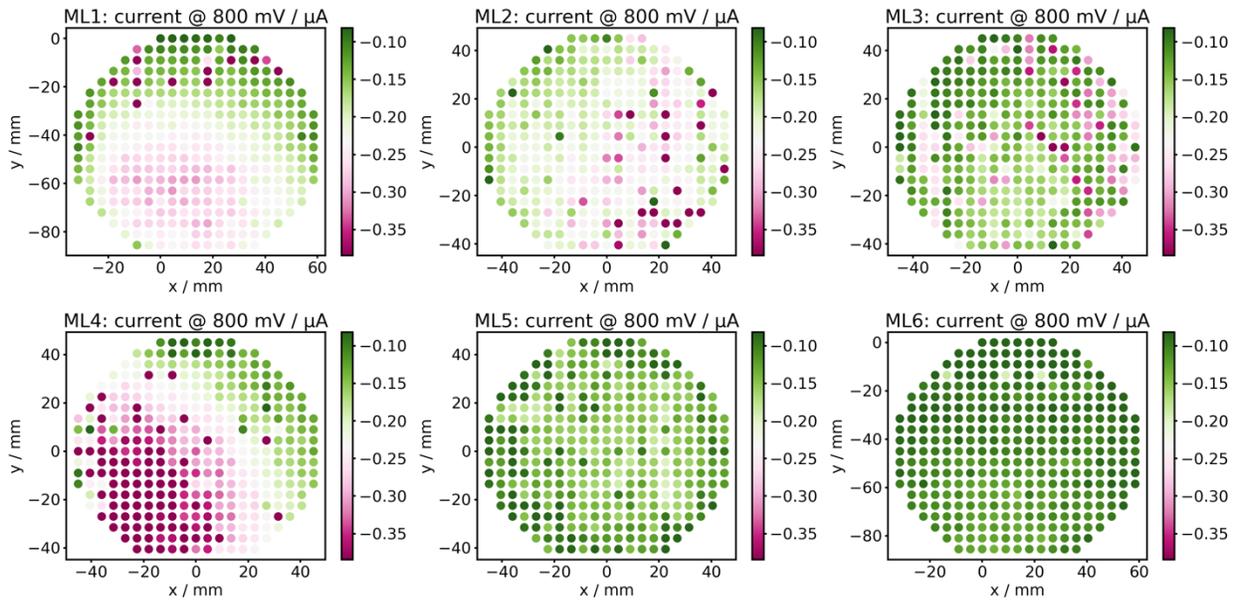

*SI 16: Activity maps for the oxygen reduction reaction at 800 mV vs RHE.*

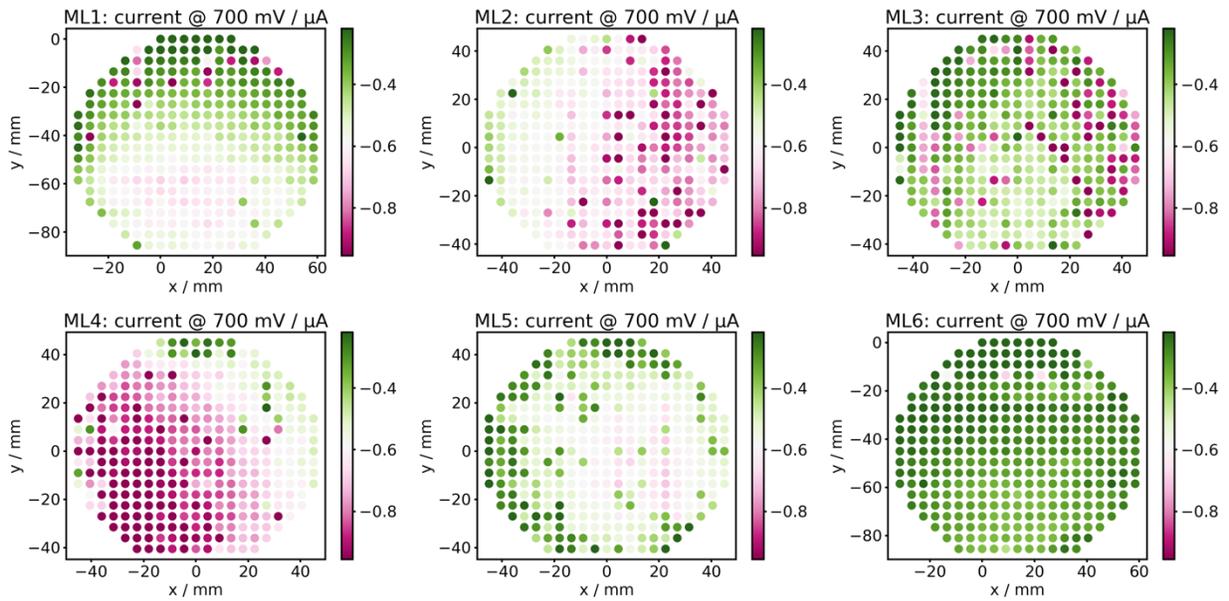

*SI 17: Activity maps for the oxygen reduction reaction at 700 mV vs RHE.*

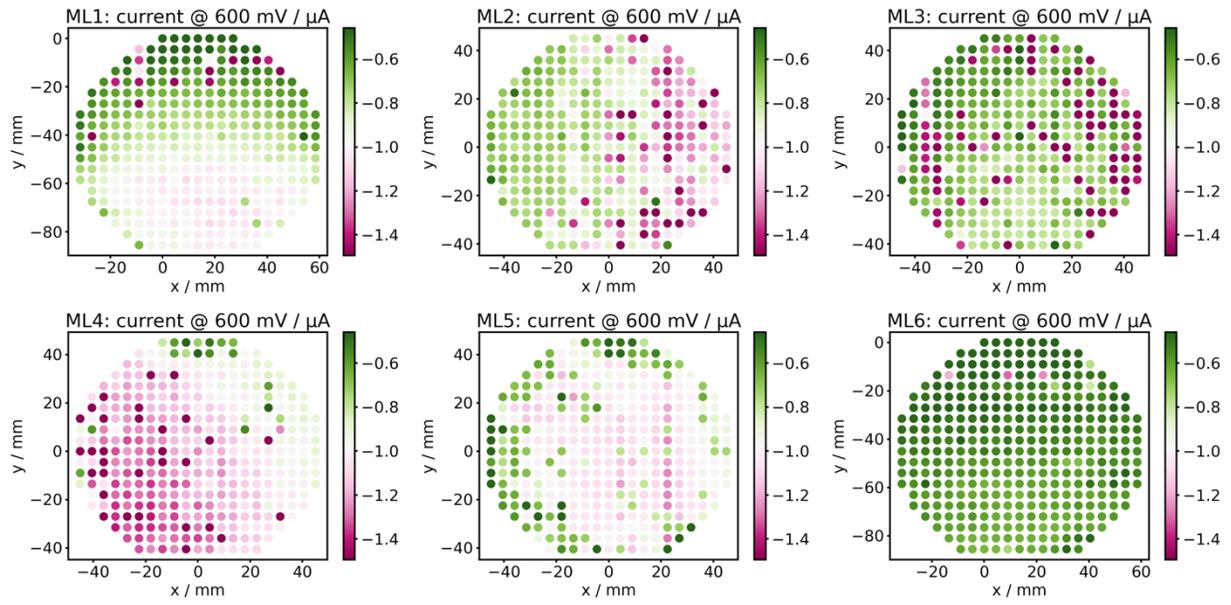

*SI 18: Activity maps for the oxygen reduction reaction at 600 mV vs RHE.*

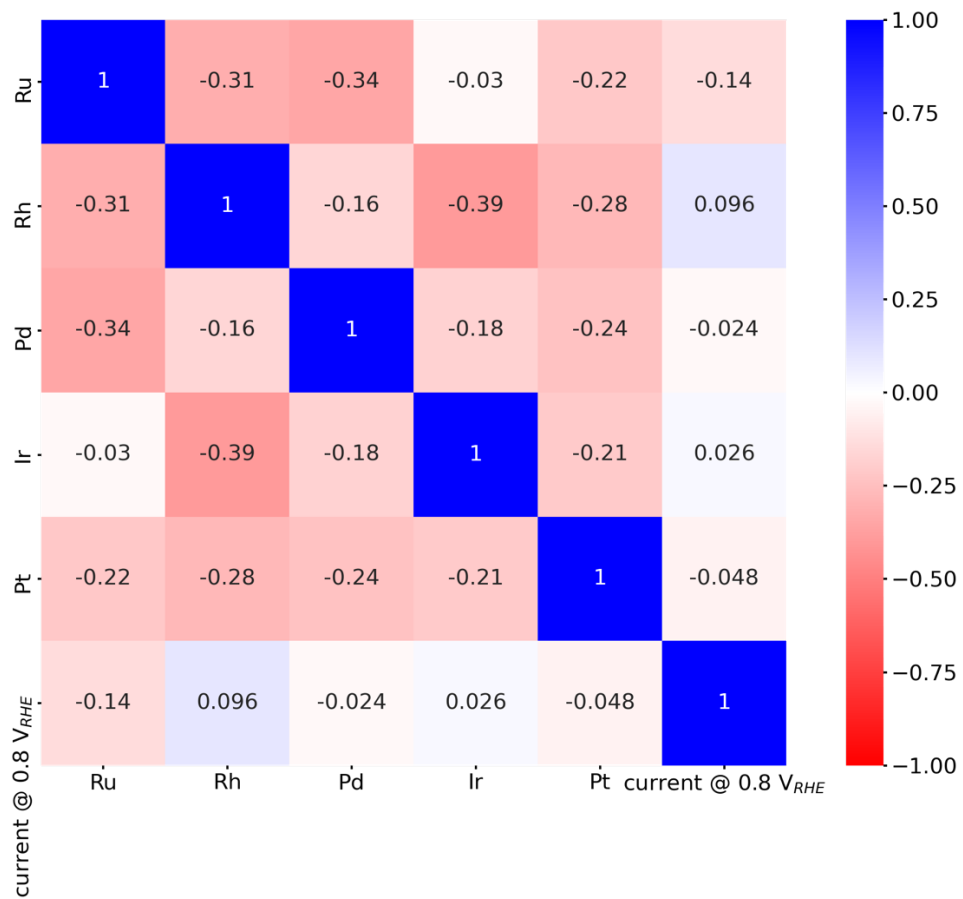

SI 19: Correlation heatmap plot (Spearman) for ML1-6.

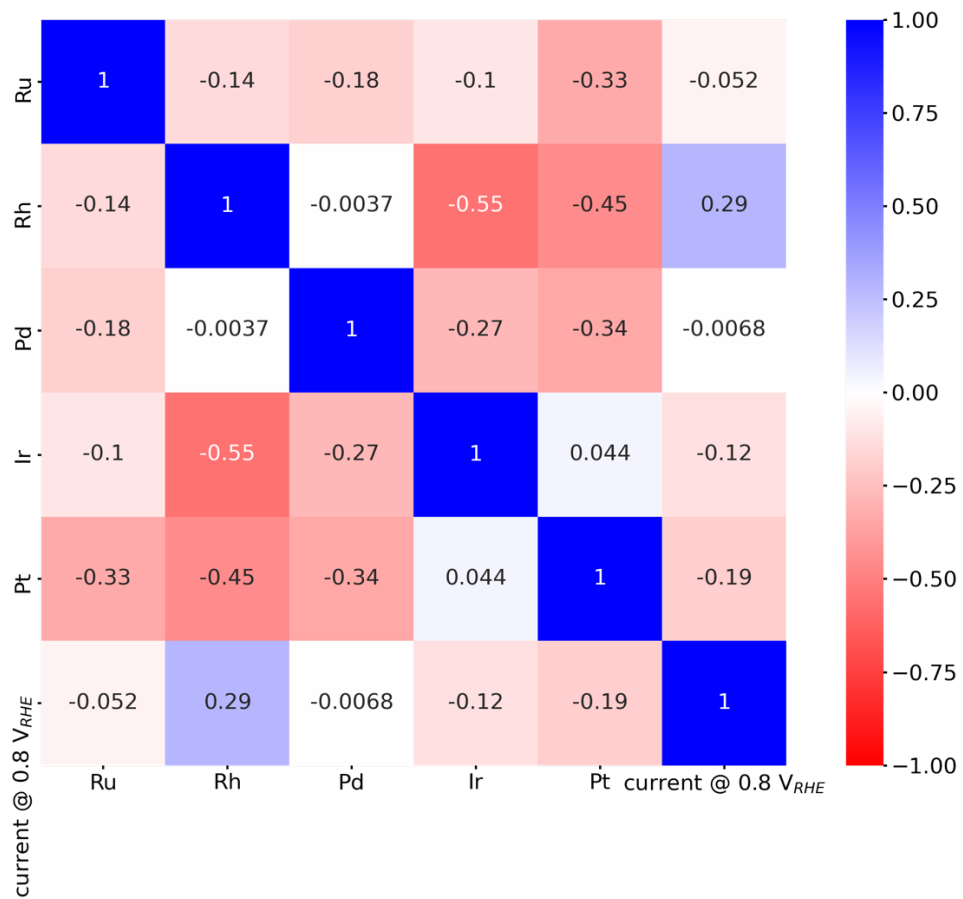

*SI 20: Correlation heatmap plot (Spearman) for the total dataset (ML1-6 + ML4b).*

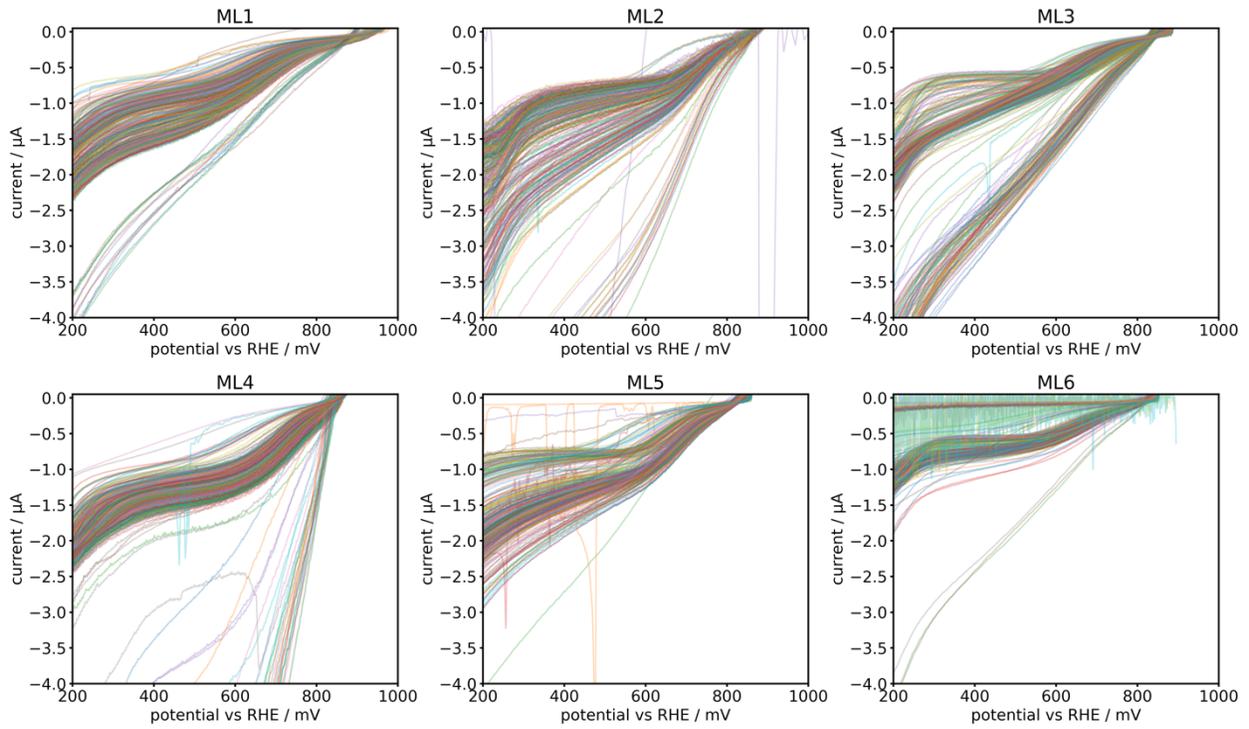

*SI 21: Linear sweep voltammograms from scanning droplet cell measurements.*

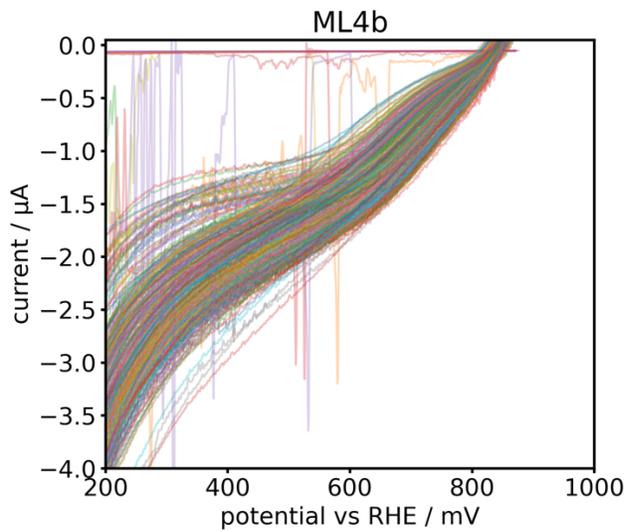

*SI 22: Linear sweep voltammograms from scanning droplet cell measurements of ML4b.*

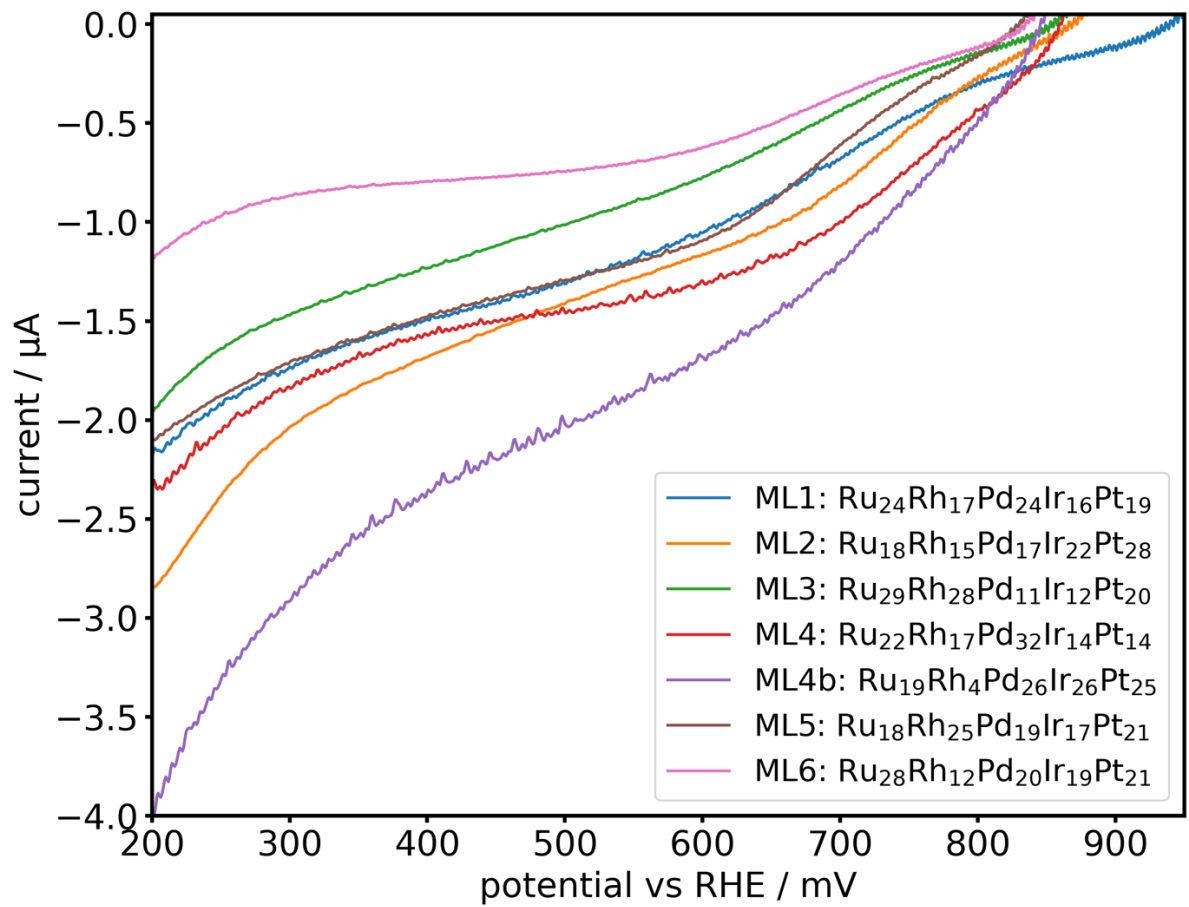

*SI 23: Linear sweep voltammograms from scanning droplet cell measurements of the most active compositions of each ML.*

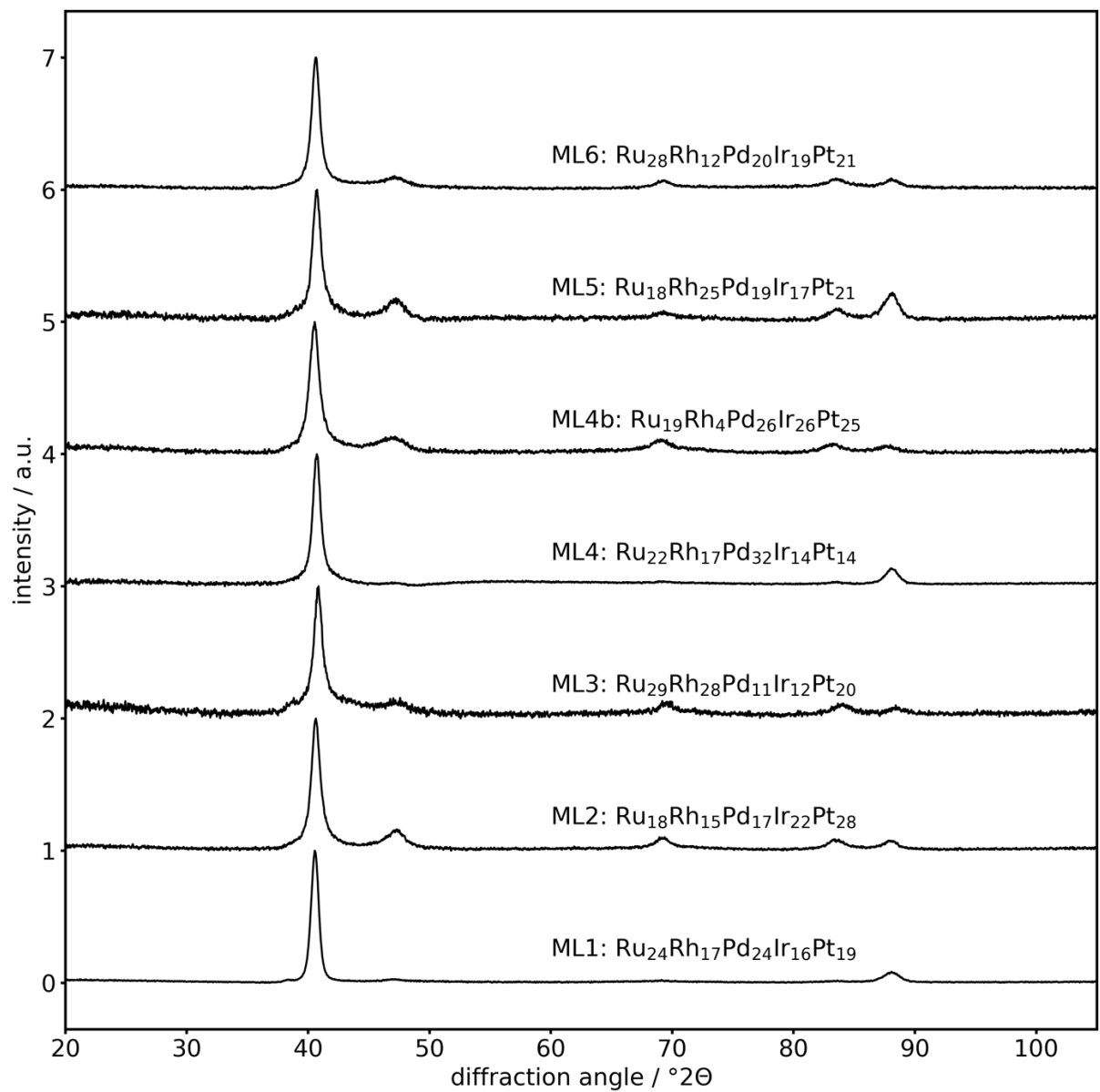

*SI 24: X-ray diffraction patterns of the most active compositions of each ML.*

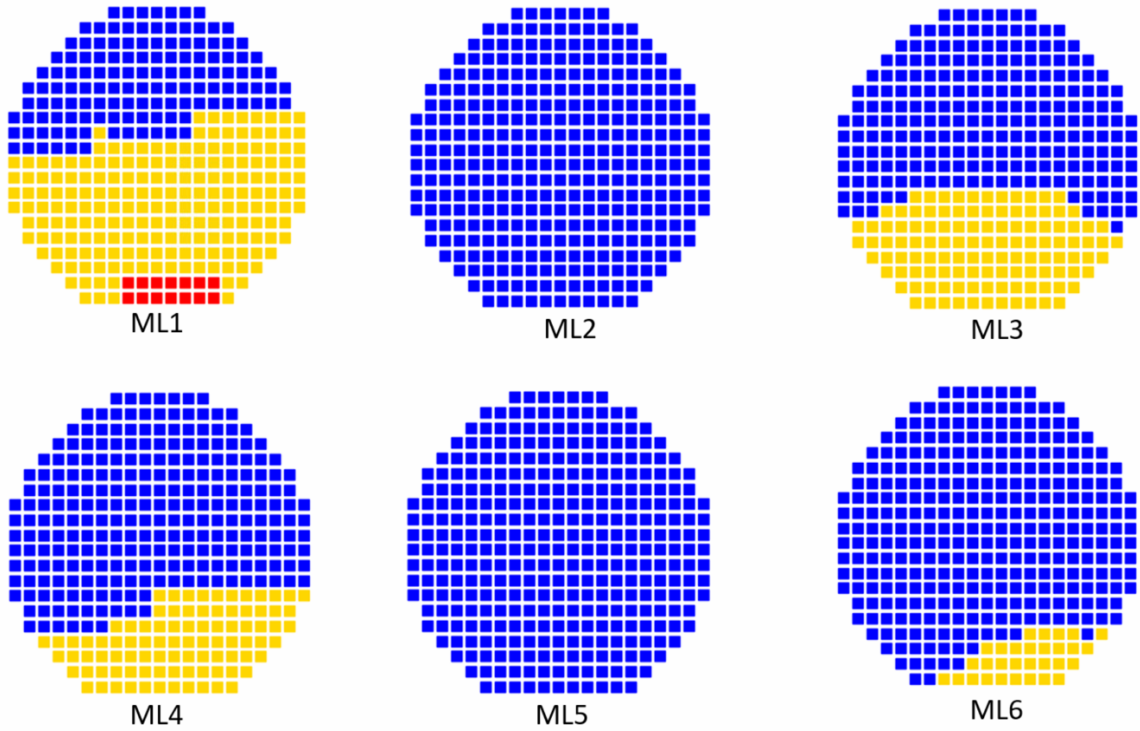

*SI 25: Results of the phase mapping. Blue: fcc, yellow: fcc + hcp phase mixture, red: hcp.*

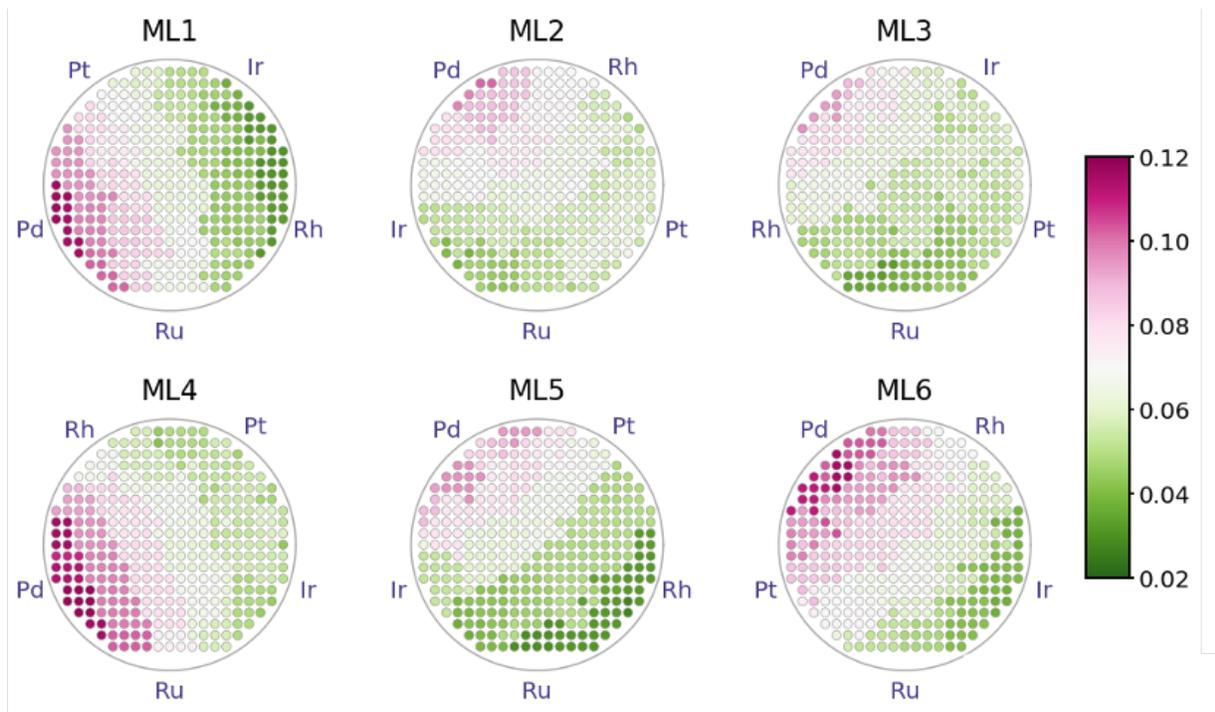

*SI 26: Results of computational model prediction for the electrochemical activity for the oxygen reduction reaction.*

# Comparison between linear regression and computational model

## LR of ML1-6 predictions for full quinary composition space

LR fitted to ML1-6 predicts a global maximum for Ru and binary Ru-Pd and a global minimum for Ir and binary Ir-Pt.

*Table 1: Top 20 compositions with the **highest** LR- predicted activity*

| Ru | Rh | Pd | Ir | Pt | current @ 0.8 VRHE |
|---|---|---|---|---|---|
| 1.00 | 0.00 | 0.00 | 0.00 | 0.00 | 0.41 |
| 0.95 | 0.00 | 0.05 | 0.00 | 0.00 | 0.41 |
| 0.90 | 0.00 | 0.10 | 0.00 | 0.00 | 0.41 |
| 0.85 | 0.00 | 0.15 | 0.00 | 0.00 | 0.41 |
| 0.80 | 0.00 | 0.20 | 0.00 | 0.00 | 0.41 |
| 0.75 | 0.00 | 0.25 | 0.00 | 0.00 | 0.41 |
| 0.70 | 0.00 | 0.30 | 0.00 | 0.00 | 0.41 |
| 0.65 | 0.00 | 0.35 | 0.00 | 0.00 | 0.40 |
| 0.60 | 0.00 | 0.40 | 0.00 | 0.00 | 0.40 |
| 0.55 | 0.00 | 0.45 | 0.00 | 0.00 | 0.40 |
| 0.50 | 0.00 | 0.50 | 0.00 | 0.00 | 0.40 |
| 0.45 | 0.00 | 0.55 | 0.00 | 0.00 | 0.40 |
| 0.40 | 0.00 | 0.60 | 0.00 | 0.00 | 0.40 |
| 0.35 | 0.00 | 0.65 | 0.00 | 0.00 | 0.40 |
| 0.95 | 0.05 | 0.00 | 0.00 | 0.00 | 0.39 |
| 0.30 | 0.00 | 0.70 | 0.00 | 0.00 | 0.39 |
| 0.95 | 0.00 | 0.00 | 0.00 | 0.05 | 0.39 |
| 0.90 | 0.05 | 0.05 | 0.00 | 0.00 | 0.39 |
| 0.25 | 0.00 | 0.75 | 0.00 | 0.00 | 0.39 |
| 0.95 | 0.00 | 0.00 | 0.05 | 0.00 | 0.39 |

*Table 2: Top 20 compositions with the **lowest** LR- predicted activity*

| Ru | Rh | Pd | Ir | Pt | current @ 0.8 VRHE |
|---|---|---|---|---|---|
| 0.00 | 0.00 | 0.00 | 1.00 | 0.00 | -0.0008 |
| 0.00 | 0.00 | 0.00 | 0.95 | 0.05 | 0.0002 |
| 0.00 | 0.00 | 0.00 | 0.90 | 0.10 | 0.0013 |
| 0.00 | 0.05 | 0.00 | 0.95 | 0.00 | 0.0015 |
| 0.00 | 0.00 | 0.00 | 0.85 | 0.15 | 0.0023 |
| 0.00 | 0.05 | 0.00 | 0.90 | 0.05 | 0.0025 |
| 0.00 | 0.00 | 0.00 | 0.80 | 0.20 | 0.0034 |
| 0.00 | 0.05 | 0.00 | 0.85 | 0.10 | 0.0036 |
| 0.00 | 0.10 | 0.00 | 0.90 | 0.00 | 0.0037 |
| 0.00 | 0.00 | 0.00 | 0.75 | 0.25 | 0.0044 |
| 0.00 | 0.05 | 0.00 | 0.80 | 0.15 | 0.0046 |
| 0.00 | 0.10 | 0.00 | 0.85 | 0.05 | 0.0048 |
| 0.00 | 0.00 | 0.00 | 0.70 | 0.30 | 0.0055 |
| 0.00 | 0.05 | 0.00 | 0.75 | 0.20 | 0.0057 |
| 0.00 | 0.10 | 0.00 | 0.80 | 0.10 | 0.0058 |
| 0.00 | 0.15 | 0.00 | 0.85 | 0.00 | 0.0060 |
| 0.00 | 0.00 | 0.00 | 0.65 | 0.35 | 0.0065 |
| 0.00 | 0.05 | 0.00 | 0.70 | 0.25 | 0.0067 |
| 0.00 | 0.10 | 0.00 | 0.75 | 0.15 | 0.0069 |
| 0.00 | 0.15 | 0.00 | 0.80 | 0.05 | 0.0071 |

**LR of ML1-6 + ML4b predictions for full quinary composition space**

LR fitted to the full dataset (ML1-6 + ML4b) predicts a global maximum for Pd and ternary Ru-Pd-Pt and a global minimum for Rh and all Rh-dominated systems.

*Table 3: Top 20 compositions with the **highest** LR- predicted activity*

| Ru | Rh | Pd | Ir | Pt | current @ 0.8 VRHE |
|---|---|---|---|---|---|
| 0.00 | 0.00 | 1.00 | 0.00 | 0.00 | 0.39 |
| 0.05 | 0.00 | 0.95 | 0.00 | 0.00 | 0.39 |
| 0.00 | 0.00 | 0.95 | 0.00 | 0.05 | 0.38 |
| 0.00 | 0.00 | 0.95 | 0.05 | 0.00 | 0.38 |
| 0.10 | 0.00 | 0.90 | 0.00 | 0.00 | 0.38 |
| 0.05 | 0.00 | 0.90 | 0.00 | 0.05 | 0.38 |
| 0.00 | 0.00 | 0.90 | 0.00 | 0.10 | 0.38 |
| 0.05 | 0.00 | 0.90 | 0.05 | 0.00 | 0.37 |
| 0.00 | 0.00 | 0.90 | 0.05 | 0.05 | 0.37 |
| 0.15 | 0.00 | 0.85 | 0.00 | 0.00 | 0.37 |
| 0.10 | 0.00 | 0.85 | 0.00 | 0.05 | 0.37 |
| 0.05 | 0.00 | 0.85 | 0.00 | 0.10 | 0.37 |
| 0.00 | 0.00 | 0.85 | 0.00 | 0.15 | 0.37 |
| 0.00 | 0.00 | 0.90 | 0.10 | 0.00 | 0.37 |
| 0.10 | 0.00 | 0.85 | 0.05 | 0.00 | 0.37 |
| 0.05 | 0.00 | 0.85 | 0.05 | 0.05 | 0.37 |
| 0.20 | 0.00 | 0.80 | 0.00 | 0.00 | 0.37 |
| 0.00 | 0.00 | 0.85 | 0.05 | 0.10 | 0.37 |
| 0.15 | 0.00 | 0.80 | 0.00 | 0.05 | 0.37 |
| 0.10 | 0.00 | 0.80 | 0.00 | 0.10 | 0.37 |

*Table 4: Top 20 compositions with the **lowest** LR- predicted activity*

| Ru | Rh | Pd | Ir | Pt | current @ 0.8 VRHE |
|---|---|---|---|---|---|
| 0.00 | 1.00 | 0.00 | 0.00 | 0.00 | -0.19 |
| 0.00 | 0.95 | 0.00 | 0.05 | 0.00 | -0.17 |
| 0.00 | 0.95 | 0.00 | 0.00 | 0.05 | -0.17 |
| 0.05 | 0.95 | 0.00 | 0.00 | 0.00 | -0.17 |
| 0.00 | 0.95 | 0.05 | 0.00 | 0.00 | -0.16 |
| 0.00 | 0.90 | 0.00 | 0.10 | 0.00 | -0.15 |
| 0.00 | 0.90 | 0.00 | 0.05 | 0.05 | -0.15 |
| 0.05 | 0.90 | 0.00 | 0.05 | 0.00 | -0.15 |
| 0.00 | 0.90 | 0.00 | 0.00 | 0.10 | -0.14 |
| 0.05 | 0.90 | 0.00 | 0.00 | 0.05 | -0.14 |
| 0.10 | 0.90 | 0.00 | 0.00 | 0.00 | -0.14 |
| 0.00 | 0.90 | 0.05 | 0.05 | 0.00 | -0.14 |
| 0.00 | 0.90 | 0.05 | 0.00 | 0.05 | -0.14 |
| 0.05 | 0.90 | 0.05 | 0.00 | 0.00 | -0.14 |
| 0.00 | 0.85 | 0.00 | 0.15 | 0.00 | -0.14 |
| 0.00 | 0.90 | 0.10 | 0.00 | 0.00 | -0.13 |
| 0.00 | 0.85 | 0.00 | 0.10 | 0.05 | -0.13 |
| 0.05 | 0.85 | 0.00 | 0.10 | 0.00 | -0.13 |
| 0.00 | 0.85 | 0.00 | 0.05 | 0.10 | -0.13 |
| 0.05 | 0.85 | 0.00 | 0.05 | 0.05 | -0.13 |

**Correlation of LR and computational model (ML1-6)**

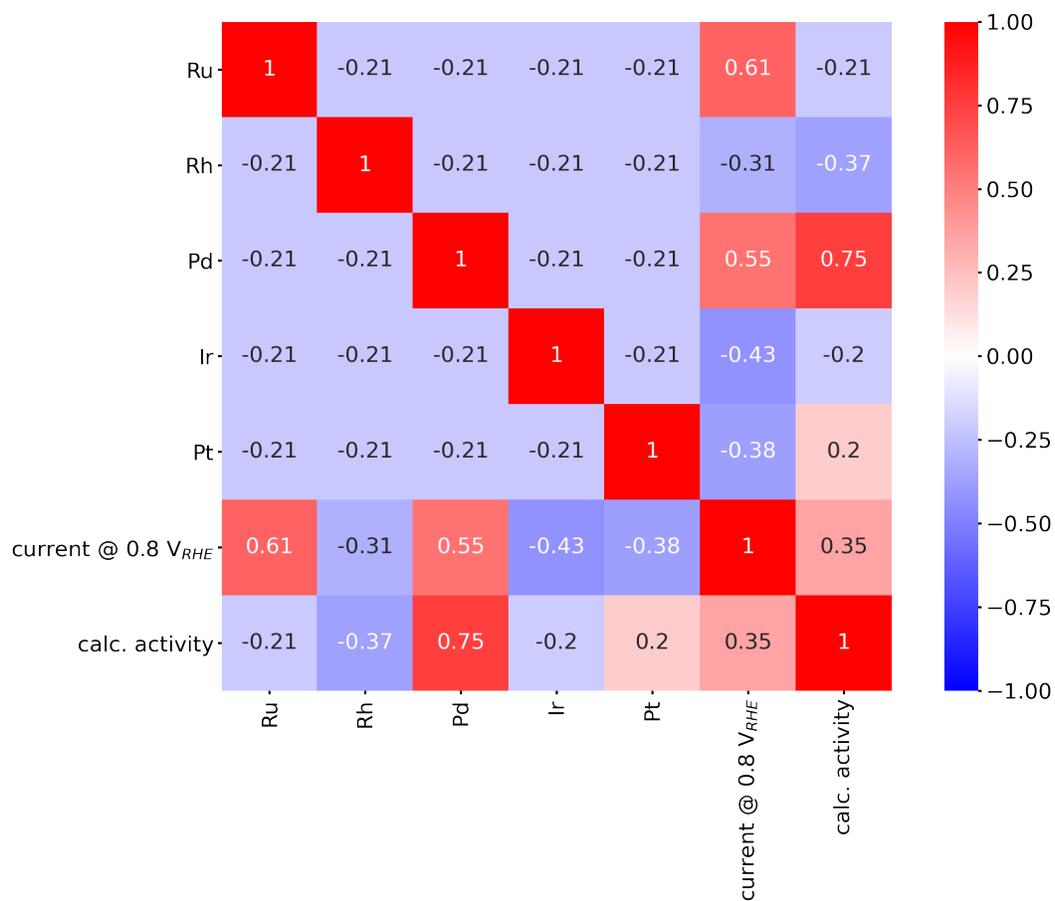

*SI 27: Correlation heatmap (Spearman correlation). Positive values indicate a positive correlation and negative values indicate a negative correlation. The label current @ 0.8 $V_{RHE}$ are the predictions by LR, calc. activity refers to the computational model.*

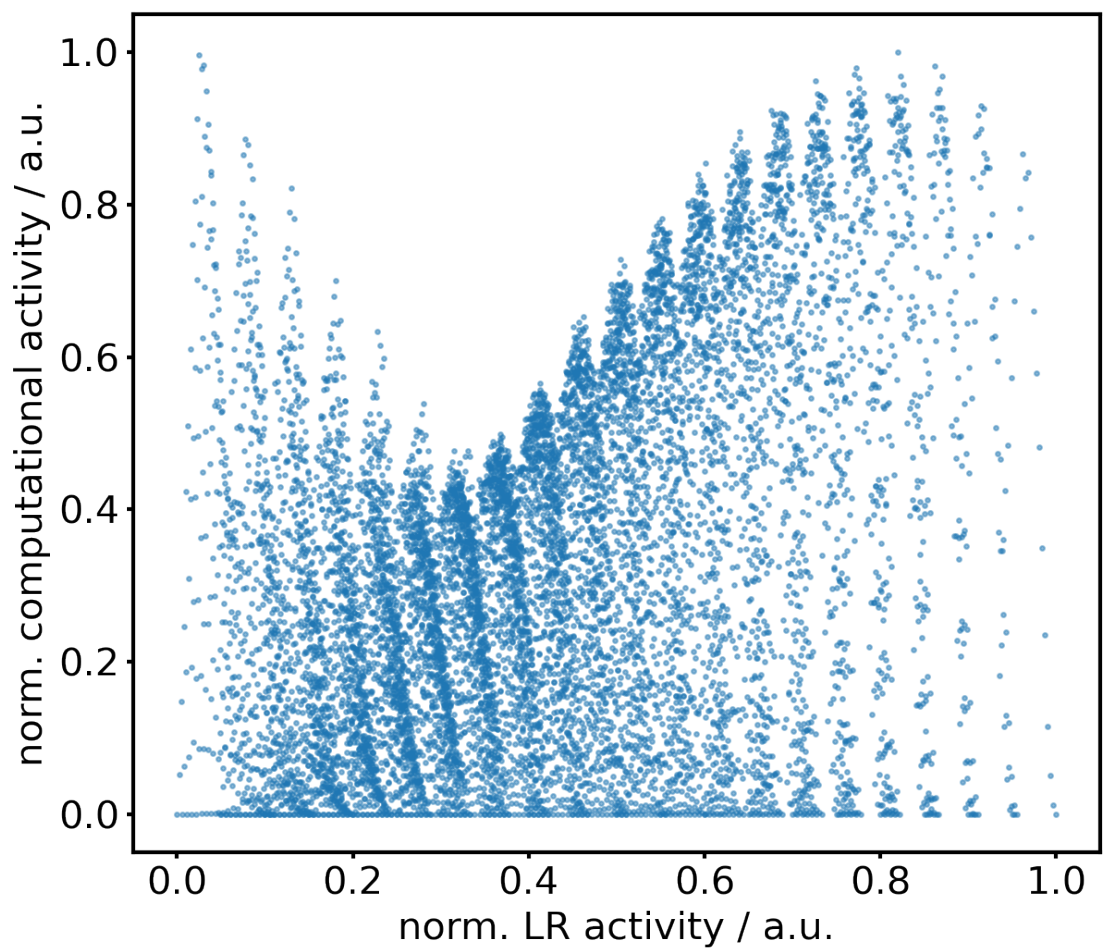

*SI 28: LR-predicted vs. computationally predicted electrochemical activity. 1 = maximum activity, 0 = minimum activity.*

## Correlation of LR and computational model (ML1-6 + ML4b)

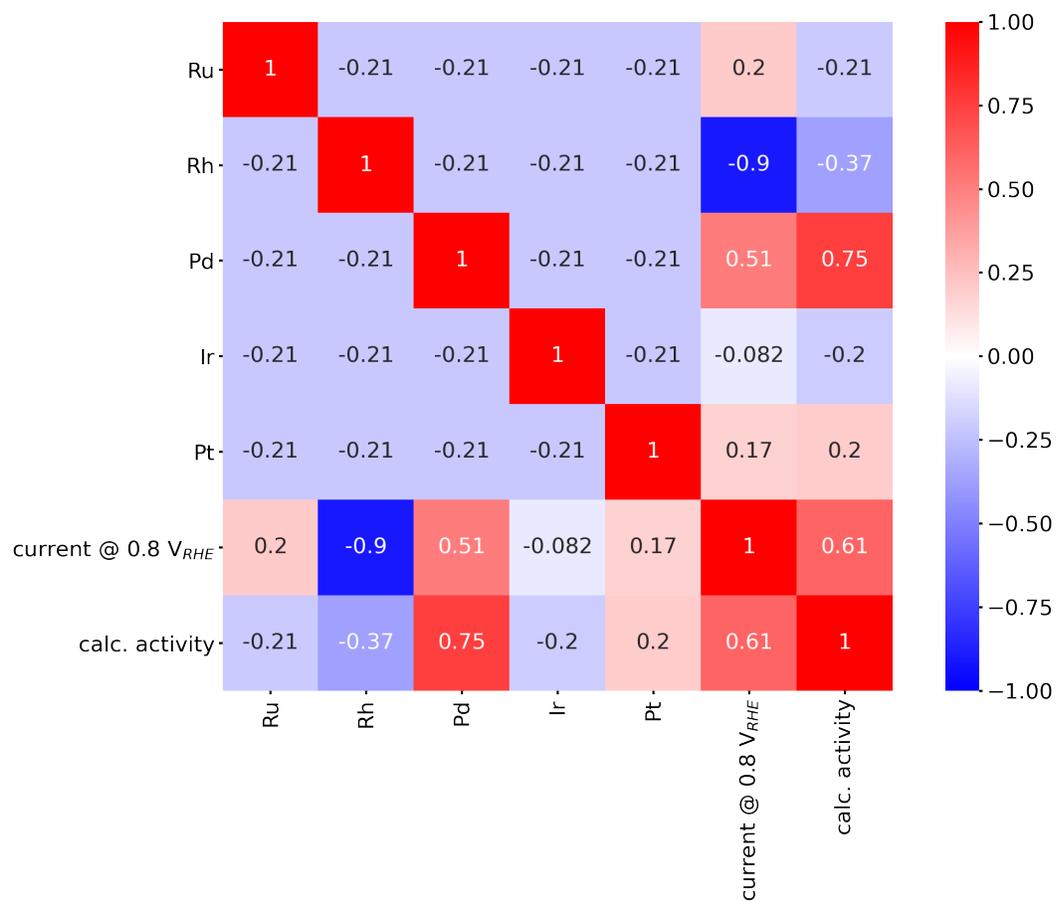

*SI 29: Correlation heatmap (Spearman correlation). Positive values indicate a positive correlation and negative values indicate a negative correlation. The label current @ 0.8 $V_{RHE}$ are the predictions by LR, calc. activity refers to the computational model.*

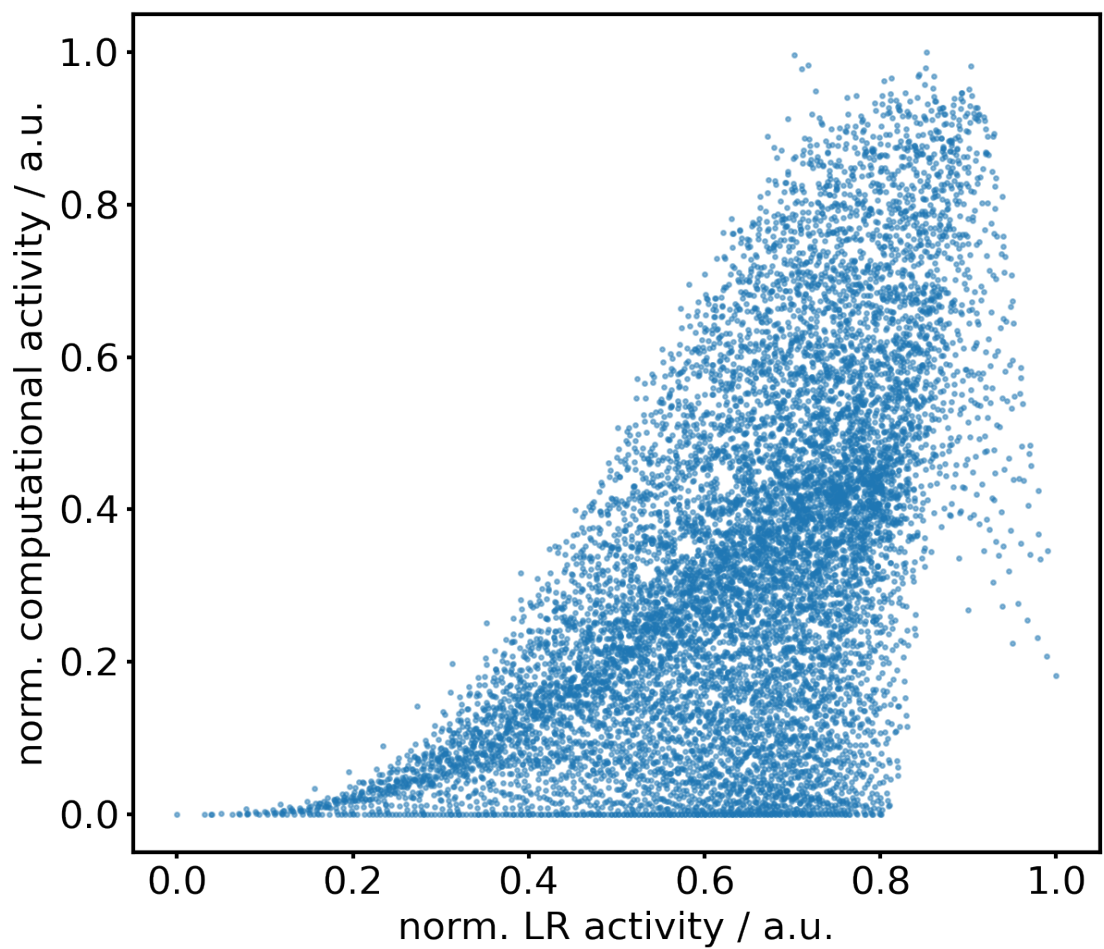

*SI 30: LR-predicted vs. computationally predicted electrochemical activity. 1 = maximum activity, 0 = minimum activity.*